%% file: main.tex
\documentclass[%
reprint,
superscriptaddress,
frontmatterverbose,
showpacs,
preprintnumbers,
nofootinbib,
nobibnotes,
amsmath,amssymb,
aps,
prc,
]{revtex4-1}
\usepackage[pdftex]{graphicx}
\usepackage{dcolumn}
\usepackage{bm}
\usepackage[pdftex]{color}
\usepackage{CJK}
\usepackage[T1]{fontenc}
\usepackage{multirow}
\def\ket#1{\left|{#1}\right\rangle}

\def\brakket#1#2#3{\left\langle{#1}\middle|{#2}\middle|{#3}\right\rangle}
\def\ve#1{{\bm{#1}}}
\def\nuc#1#2#3{{}^{#2}_{#3}\mathrm{#1}}
\def\urm#1{\scriptstyle{\text{\textrm{\textmd{\textup{#1}}}}}}
\def\uurm#1{\scriptscriptstyle{\text{\textrm{\textmd{\textup{#1}}}}}}
\def\avr#1{\left\langle{#1}\right\rangle}

\def\ca#1{{\mathcal{#1}}}
\let\temp\epsilon
\let\epsilon\varepsilon
\let\varepsilon\temp
\let\temp\relax
\let\temp\phi
\let\phi\varphi
\let\varphi\temp
\let\temp\relax
\DeclareMathOperator{\laplace}{\Delta}
\DeclareMathOperator{\sgn}{sgn}
\begin{document}
% 
%%%%%%%%%%%%%%%%%%%%%%%%%%%%%%%%%%%%%%%%%%%%%%%%%% 
\begin{CJK*}{UTF8}{}
  \preprint{RIKEN-QHP-481}
  \preprint{RIKEN-iTHEMS-Report-21}
  \title{
    Second and fourth moments of the charge density and neutron-skin thickness of atomic nuclei}
  \author{Tomoya Naito (\CJKfamily{min}{内藤智也})}
  \email{
    tomoya.naito@phys.s.u-tokyo.ac.jp}
  \affiliation{
    Department of Physics, Graduate School of Science, The University of Tokyo,
    Tokyo 113-0033, Japan}
  \affiliation{
    RIKEN Nishina Center, Wako 351-0198, Japan}
  \author{Gianluca Col\`{o}}
  \email{
    colo@mi.infn.it}
  \affiliation{
    Dipartimento di Fisica, Universit\`{a} degli Studi di Milano,
    Via Celoria 16, 20133 Milano, Italy}
  \affiliation{
    INFN, Sezione di Milano,
    Via Celoria 16, 20133 Milano, Italy}
  \author{Haozhao Liang (\CJKfamily{gbsn}{梁豪兆})}
  \email{
    haozhao.liang@phys.s.u-tokyo.ac.jp}
  \affiliation{
    Department of Physics, Graduate School of Science, The University of Tokyo,
    Tokyo 113-0033, Japan}
  \affiliation{
    RIKEN Nishina Center, Wako 351-0198, Japan}
  \author{Xavier Roca-Maza}
  \email{
    xavier.roca.maza@mi.infn.it}
  \affiliation{
    Dipartimento di Fisica, Universit\`{a} degli Studi di Milano,
    Via Celoria 16, 20133 Milano, Italy}
  \affiliation{
    INFN, Sezione di Milano,
    Via Celoria 16, 20133 Milano, Italy}
  \date{\today}
  %%%%%%%%%%%%%%%%%%%%%%%%%%%%%%%%%%%%%%%%%%%%%%%%%% 
  \begin{abstract}
    A method is presented to extract the neutron-skin thickness of atomic nuclei from the second and fourth moments of the electric charge distribution.
    We show that the value of the proton fourth moment must be independently known in order to estimate the neutron skin thickness experimentally.
    To overcome this problem,
    we propose the use of a strong linear correlation among the second and fourth moments of the proton distribution as calculated with several energy density functionals of common use.
    We take special care in estimating the errors associated with the different contributions to the neutron radius and show,
    for the first time, 
    the analytic expressions for the spin-orbit contribution to the charge fourth moments of neutrons and protons.
    To reduce the uncertainty on the extraction of the neutron radius, two neighboring even-even isotopes are used.
    Nevertheless, the error on the fourth moment of the proton distribution, even if determined or assumed with large accuracy, dominates and prevents the present method from being applied for a sound determination of the neutron skin thickness. 
  \end{abstract}
  \maketitle
\end{CJK*}
%%%%%%%%%%%%%%%%%%%%%%%%%%%%%%%%%%%%%%%%%%%%%%%%%% 
% 
% Introduction
\input{introduction}
% 
% Calculation
\input{calculation}
% 
% Conclusion
\input{conclusion}
% 
% Acknowledgement
\input{acknowledgement}
\appendix
% Appendix
\input{appendix_detail}
\end{document}

%% file: introduction.tex
% -*- coding: utf-8 -*-
%%%%%%%%%%%%%%%%%%%%%%%%%%%%%%%%%%%%%%%%%%%%%%%%%% 
% 
% 4th moment paper manuscript 
% Introduction Part
% 
% Tomoya Naito, Gianluca Colo',
% Haozhao Liang, and Xavier Roca-Maza
% 
%%%%%%%%%%%%%%%%%%%%%%%%%%%%%%%%%%%%%%%%%%%%%%%%%% 
% 
\section{Introduction}
\par
The study of the neutron-skin thickness $ \Delta r_{np} = \sqrt{\avr{r^2}_n} - \sqrt{\avr{r^2}_p} $ 
of atomic nuclei has become one of the hottest topics in nuclear physics during the past decades~\cite{
  Bender2003Rev.Mod.Phys.75_121,
  Tsang2012Phys.Rev.C86_015803,
  Lattimer2012Annu.Rev.Nucl.Part.Sci.62_485,
  Hebeler2015Annu.Rev.Nucl.Part.Sci.65_457,
  Thiel2019J.Phys.G46_093003}.
Here, $ \avr{r^2}_p $ and $ \avr{r^2}_n $ denote the second moments of the proton and neutron density distributions,
$ \rho_p $ and $ \rho_n $, respectively.
A precise determination of the neutron-skin thickness of a heavy nucleus sets a basic constraint on the nuclear symmetry energy,
in particular,
its density dependence around the saturation density~\cite{
  Yoshida2004Phys.Rev.C69_024318,
  Yoshida2006Phys.Rev.C73_044320,
  Roca-Maza2013Phys.Rev.C87_034301,
  Vinas2014Eur.Phys.J.A50_27,
  Typel2014Phys.Rev.C89_064321,
  Pais2016Phys.Rev.C93_045802,
  Mondal2016Phys.Rev.C93_064303}.
For example, the neutron-skin thickness of $ \nuc{Pb}{208}{} $ is known to be directly related to the slope parameter of the symmetry energy $ L $ by
$ \Delta r_{np} \left( \nuc{Pb}{208}{} \right) \left[ \mathrm{fm} \right] = 0.101 (3) + 0.001 \, 47 (5) L \left[ \mathrm{MeV} \right] $ ($ r = 0.98 $)
if one exploits the prediction by
a large and representative set of modern nuclear energy density functionals (EDFs)~\cite{
  Roca-Maza2011Phys.Rev.Lett.106_252501}.
The equation of state of nuclear matter, which provides the value of $ L $,
is also known to be related to a wide range of questions in nuclear physics and astrophysics~\cite{
  Myers1980Nucl.Phys.A336_267,
  Stringari1983Prog.Theor.Phys.Suppl.74_367,
  Sagawa2002Phys.Rev.C65_064314,
  Chen2003Phys.Rev.Lett.90_162701,
  Baran2005Phys.Rep.410_335,
  Li2008Phys.Rep.464_113,
  DiToro2010J.Phys.G37_083101,
  Gandolfi2012Phys.Rev.C85_032801,
  Colo2014Eur.Phys.J.A50_26,
  Hebeler2013AstrophysJ.773_11,
  Malik2018Phys.Rev.C98_035804,
  Horowitz2019Ann.Phys.411_167992,
  Morfouace2019Phys.Lett.B799_135045,
  Burrello2019Front.Phys.7_53,
  Tong2020Phys.Rev.C101_035802}.
For more detail, see the review papers, e.g.,~Refs.~\cite{
  Lattimer2012Annu.Rev.Nucl.Part.Sci.62_485,
  Vinas2014Eur.Phys.J.A50_27,
  Lattimer2014Nucl.Phys.A928_276,
  Baldo2016Prog.Part.Nucl.Phys.91_203,
  Gandolfi2019J.Phys.G46_103001}.
Yet, our knowledge of neutron-skin thickness is limited even in the stable nuclei,
and neutron-skin thickness of the unstable nuclei has not been measured yet.
\par
The parity-violating elastic electron scattering~\cite{
  Roca-Maza2011Phys.Rev.Lett.106_252501,
  Donnelly1990Prog.Part.Nucl.Phys.24_179,
  Paschke2011_,
  Armstrong2012Annu.Rev.Nucl.Part.Sci.62_337,
  Horowitz2012Phys.Rev.C85_032501,
  Abrahamyan2012Phys.Rev.Lett.108_112502,
  Souder2016Front.Phys.11_111301,
  Adhikari2021Phys.Rev.Lett.126_172502}
and the isotopic ratio of atomic parity violation~\cite{
  Fortson1990Phys.Rev.Lett.65_2857}
were suggested as clean and model-independent probes of neutron densities.
However, measuring parity-violating asymmetries of the order of a part per $ 10^6 $ is challenging.
The present result is
$ \Delta r_{np} = 0.283 \pm 0.071 \, \mathrm{fm} $ at the PREX-II experiment~\cite{
  Paschke2011_,
  Adhikari2021Phys.Rev.Lett.126_172502}
for $ \nuc{Pb}{208}{} $.
The ambitious efforts in JLab aim at determining the $ \Delta r_{np} $ of $ \nuc{Ca}{48}{} $ with higher precisions as well~\cite{
  Horowitz2014Eur.Phys.J.A50_48}.
\par
The hadronic probes, 
including polarized-proton scattering~\cite{
  Zenihiro2010Phys.Rev.C82_044611,
  Sakaguchi2017Prog.Part.Nucl.Phys.97_1},
$ \alpha $ scattering~\cite{
  Tatischeff1972Phys.Rev.C5_234},
anti-protonic atoms~\cite{
  AlexBrown2007Phys.Rev.C76_034305},
$ \pi^{\pm} $ scattering~\cite{
  Johnson1979Phys.Rev.Lett.43_844,
  Barnett1985Phys.Lett.B156_172},
and 
anti-proton scattering~\cite{
  Lenske2007Phys.Lett.B647_82,
  Makiguchi2020Phys.Rev.C102_034614},
as well as the nuclear excitations, such as isovector resonances~\cite{
  Krasznahorkay1999Phys.Rev.Lett.82_3216},
have been also widely used or proposed to determine the neutron-skin thickness and cover
a large area of the nuclear chart.
Nevertheless, even if some of these experiments reach small errors, all hadronic probes require model assumptions to deal with the strong force,
which, in principle, introduces systematic uncertainties.
\par
In contrast, the study of the charge density distribution $ \rho_{\urm{ch}} $ of atomic nuclei,
which is essentially dominated by the proton density distribution $ \rho_p $,
can be experimentally determined with no model dependence via elastic electron scattering~\cite{
  Lyman1951Phys.Rev.84_626,
  Hofstadter1953Phys.Rev.91_422,
  Pidd1953Phys.Rev.92_436,
  Hofstadter1953Phys.Rev.92_978,
  Hofstadter1954Phys.Rev.95_512,
  Hofstadter1956Rev.Mod.Phys.28_214}.
The $ \rho_{\urm{ch}} $ of many stable nuclei has been measured with very high accuracy~\cite{
  DeVries1987At.DataNucl.DataTables36_495}.
As a big step further, the electron scattering of unstable nuclei is foreseen in the near future,
for instance, in the SCRIT facility in RIKEN~\cite{
  Wakasugi2013Nucl.Instrum.MethodsPhys.Res.Sect.B317_668,
  Wakasugi2004Nucl.Instrum.MethodsPhys.Res.Sect.A532_216,
  Tsukada2017Phys.Rev.Lett.118_262501}
and
in the ELISe facility in FAIR~\cite{
  Antonov2011Nucl.Instrum.MethodsPhys.Res.Sect.A637_60,
  Berg2011Nucl.Instrum.MethodsPhys.Res.Sect.A640_123}.
Nowadays, essentially the only way to provide the information of the charge radii of unstable nuclei is the laser spectroscopy of atoms.
The laser spectroscopy of atoms was established in the late 1910s~\cite{
  Aronberg1918Astrophys.J.47_96,
  Merton1920Proc.R.Soc.Lond.A96_388}
and has been applied to long-lived unstable nuclei since the 1960s~\cite{
  Marrus1965Phys.Rev.Lett.15_813,
  Huehnermann1966Phys.Lett.21_303,
  Jacquinot1979Rep.Prog.Phys.42_773,
  Kluge2003Spectrochim.ActaPartB58_1031,
  Campbell2016Prog.Part.Nucl.Phys.86_127,
  Flambaum2019Phys.Rev.A100_032511}.
The charge radii of nuclei located on wide region of the nuclear chart have been measured~\cite{
  Angeli2013At.DataNucl.DataTables99_69},
and are still being measured at many radioactive isotope beam facilities.
\par
Recently, the fourth moment of the charge distribution has been highlighted as a possible proxy to access information of the neutron root-mean-square radius~\cite{
  Kurasawa2019Prog.Theor.Exp.Phys.2019_113D01,
  Reinhard2020Phys.Rev.C101_021301,
  Kurasawa2021Prog.Theor.Exp.Phys.2021_013D02}.
Kurasawa and Suzuki~\cite{
  Kurasawa2019Prog.Theor.Exp.Phys.2019_113D01}
suggested that the fourth moment of the charge density distribution $ \avr{r^4}_{\urm{ch}} $,
which can be measured by the electron scattering~\cite{
  Kurasawa2021Prog.Theor.Exp.Phys.2021_013D02}
or the laser spectroscopy~\cite{
  Papoulia2016Phys.Rev.A94_042502},
includes the information of the neutron radius and thus the neutron-skin thickness.
This is because the neutron distributions $ \rho_n $ of atomic nuclei do contribute to their charge density distributions $ \rho_{\urm{ch}} $~\cite{
  Bertozzi1972Phys.Lett.B41_408,
  Kurasawa2000Phys.Rev.C62_054303,
  Naito2020Phys.Rev.C101_064311}
since a neutron has a finite size and has a corresponding internal charge distribution, which is usually encoded in the electromagnetic form factor of the neutron.
In other words, precise measurements of $ \rho_{\urm{ch}} $ may be able to provide information on $ \rho_n $ as well as $ \rho_p $
and, thus, determine the neutron-skin thickness $ \Delta r_{np} $.
For instance, Ref.~\cite{
  Kurasawa2021Prog.Theor.Exp.Phys.2021_013D02}
showed the feasibility to extract $ \avr{r^2}_n $ using $ \avr{r^2}_{\urm{ch}} $ and $ \avr{r^4}_{\urm{ch}} $ for $ \nuc{Ca}{40}{} $, $ \nuc{Ca}{48}{} $, and $ \nuc{Pb}{208}{} $ isotopes
by using the linear correlations among second and fourth moments of proton, neutron, and charge density distributions,
and eventually, the uncertainty of $ \avr{r^2}_n $ is quite small.
Indeed, this relied on a correlation for these specific nuclei based on a specific type of models,
and, hence, it is questionable whether that method can be applied, in general.
\par
To answer this question,
in this paper, we discuss the feasibility of extracting $ \avr{r^2}_n $ from
the second and fourth moments of the charge density distribution
$ \avr{r^2}_{\urm{ch}} $ and $ \avr{r^4}_{\urm{ch}} $,
applying the general modeling of electromagnetic form factors of both protons and neutrons
avoiding as much as possible the use of model-induced correlations.
We also explore a method to extract the neutron-skin thickness by employing the information of $ \rho_{\urm{ch}} $ of two neighboring even-even isotopes
to cancel large part of the spin-orbit contributions to $ \avr{r^2}_{\urm{ch}} $ and $ \avr{r^4}_{\urm{ch}} $
and reduce the uncertainty due to the nucleon form factors and the pairing correlation.
To extract $ \avr{r^2}_n $ from $ \avr{r^2}_{\urm{ch}} $ and $ \avr{r^4}_{\urm{ch}} $,
we will show that the key issue is how to accurately determine $ \avr{r^4}_p $.
\par
This paper is organized as follows:
First, the general equations for $ \avr{r^2}_{\urm{ch}} $ and $ \avr{r^4}_{\urm{ch}} $ will be given in Sec.~\ref{sec:general}
as functions of the second and fourth moments of the neutron and proton density distributions and of the parameters defining the neutron and proton electric form factors.  
Second, a novel equation will be introduced in Sec.~\ref{sec:diff}
to reduce the uncertainty due to the magnetic contribution and nucleon form factors.
In this equation,
two neighboring even-even nuclei are used
in which the same single-particle orbitals are being filled and, thus,
the uncertainties associated with the latter effects are expected to be reduced. 
Third, the possibility to derive theoretically the fourth moment of the proton distribution $ \avr{r^4}_p $ will be discussed in Sec.~\ref{sec:R4p}.
Then, we will show the benchmark calculation of the novel method in Sec.~\ref{sec:isotope_shift}.
We will also show the uncertainty due to the nucleon form factors in Sec.~\ref{sec:form_factor}.
Finally, the conclusion and perspectives will be given in Sec.~\ref{sec:conc}.

%% file: calculation.tex
% -*- coding: utf-8 -*-
%%%%%%%%%%%%%%%%%%%%%%%%%%%%%%%%%%%%%%%%%%%%%%%%%% 
% 
% 4th moment paper manuscript 
% Calculation Part
% 
% Tomoya Naito, Gianluca Colo',
% Haozhao Liang, and Xavier Roca-Maza
% 
%%%%%%%%%%%%%%%%%%%%%%%%%%%%%%%%%%%%%%%%%%%%%%%%%% 
% 
\section{Second and Fourth Moments of Charge Distribution}
\label{sec:general}
\par
First, we would recall the relationship between $ \avr{r^n}_{\urm{ch}} $ and $ \avr{r^n}_{\tau} $,
which is originally derived in Refs.~\cite{
  Kurasawa2019Prog.Theor.Exp.Phys.2019_113D01,
  Reinhard2021Phys.Rev.C103_054310}.
It is convenient to consider the finite-size effects of nucleons on the charge density distribution in the momentum space, i.e.,
\begin{widetext}
  \begin{equation}
    \label{eq:charge}
    \tilde{\rho}_{\urm{ch}} \left( q \right)
    =
    \sum_{\tau = p, \, n}
    \left[
      \tilde{G}_{\urm{E} \tau} \left( q^2 \right)
      \tilde{\rho}_{\tau} \left( q \right)
      +
      \frac{\tilde{G}_{\urm{M} \tau} \left( q^2 \right) - \tilde{G}_{\urm{E} \tau} \left( q^2 \right)}{1 + q^2/4M_{\tau}^2}
      \left(
        \frac{q^2}{4M_{\tau}^2}
        \tilde{\rho}_{\tau} \left( q \right)
        +
        \frac{q}{2M_{\tau}}
        \tilde{F}_{\urm{T} \tau} \left( q \right)
      \right)
    \right],
  \end{equation}
\end{widetext}
where
$ M_{\tau} $ is the nucleon mass~\cite{
  Zyla2020Prog.Theor.Exp.Phys.2020_083C01},
$ \tilde{G}_{\urm{E} \tau} $ and $ \tilde{G}_{\urm{M} \tau} $ are the electric and magnetic form factors of nucleons, 
$ \rho_{\urm{ch}} $, $ \rho_p $, and $ \rho_n $, respectively, are
the charge, proton, and neutron density distributions, 
which are assumed to have spherical symmetry in this paper,
$ \tilde{\rho} $ is the Fourier transform of the density $ \rho $,
and
$ \tilde{F}_{\urm{T} \tau} $ is the tensor form factor,
whose definition is given in Eq.~\eqref{eq:form_T2}.
Note that $ \tilde{\rho}_{\urm{ch}} $ is sometimes called the (charge) form factor of the nucleus.
The Fourier transform is defined by 
\begin{equation}
  \label{eq:Fourier}
  \tilde{\rho}_{\tau} \left( q \right)
  =
  \int
  \rho_{\tau} \left( r \right) \,
  e^{- i \ve{q} \cdot \ve{r}}
  \, d \ve{r}
  =
  4 \pi
  \int_0^{\infty}
  \rho_{\tau} \left( r \right) \,
  \frac{\sin \left( q r \right)}{qr}
  r^2 \, dr.
\end{equation}
\par
The $ 2n $th moment of $ \rho_{\tau} $ is defined by
\begin{equation}
  \avr{r^{2n}}_{\tau}
  =
  \frac{\int \rho_{\tau} \left( \ve{r} \right) r^{2n} \, d \ve{r}}
  {\int \rho_{\tau} \left( \ve{r} \right) \, d \ve{r}}.
\end{equation}
In particular, with the assumption of spherical symmetry, this expression for $ \rho_{\urm{ch}} $ can be simplified into~\footnote{
  Note that the expression of $ \rho_{\urm{ch}} $ without assuming the spherical symmetry has been shown recently in Ref.~\cite{
    Reinhard2021Phys.Rev.C103_054310}.}
\begin{widetext}
  \begin{equation}
    \label{eq:R2n_simple}
    \avr{r^{2n}}_{\urm{ch}}
    =
    \frac{\left( -1 \right)^n 4 \pi}{Z}
    \sum_{\tau = p, \, n}
    \int_0^{\infty} 
    \left\{
      \frac{1}{q}
      \frac{d^{2n}}{dq^{2n}}
      \tilde{G}_{\urm{E} \tau} \left( q^2 \right) \,
      \sin \left( q r \right)
    \right\}_{q = 0}
    \rho_{\tau} \left( r \right) \,
    r \, dr,
  \end{equation}
  where derivation of this equation is shown in Appendix~\ref{sec:appen_detail}.
  Using Eq.~\eqref{eq:R2n_simple}, the second and fourth moments of $ \rho_{\urm{ch}} $ read
  \begin{subequations}
    \begin{align}
      \avr{r^2}_{\urm{ch}}
      & =
        \avr{r^2}_p
        +
        \left(
        r_{\urm{E} p}^2
        +
        \frac{N}{Z}
        r_{\urm{E} n}^2
        \right)
        +
        \avr{r^2}_{\urm{SO} p}
        +
        \frac{N}{Z}
        \avr{r^2}_{\urm{SO} n}, 
        \label{eq:second} \\
      \avr{r^4}_{\urm{ch}}
      & =
        \avr{r^4}_p
        +
        \frac{10}{3}
        \left(
        r_{\urm{E} p}^2
        \avr{r^2}_p
        +
        \frac{N}{Z}
        r_{\urm{E} n}^2
        \avr{r^2}_n
        \right)
        +
        \left(
        r_{\urm{E} p}^4
        +
        \frac{N}{Z}
        r_{\urm{E} n}^4
        \right)
        +
        \avr{r^4}_{\urm{SO} p}
        +
        \frac{N}{Z}
        \avr{r^4}_{\urm{SO} n},
        \label{eq:fourth}
    \end{align}
  \end{subequations}
  where
  $ r_{\urm{E} \tau}^2 $ and $ r_{\urm{E} \tau}^4 $ are the second and fourth moments of charge distribution of the nucleon $ \tau $, respectively
  [note that $ r_{\urm{E} \tau}^4 \ne \left( r_{\urm{E} \tau}^2 \right)^2 $].
  For a detailed derivation, see Appendix~\ref{sec:appen_charge}.
  Here, $ \avr{r^n}_{\urm{SO} \tau} $ is called the spin-orbit contribution due to the existence of the magnetic form factor,
  and if one considers only the first term of Eq.~\eqref{eq:charge},
  it vanishes.
  The spin-orbit contributions $ \avr{r^n}_{\urm{SO} \tau} $ read
  \begin{subequations}
    \begin{align}
      \avr{r^2}_{\urm{SO} \tau}
      & \simeq
        \frac{\kappa_{\tau}}{M_{\tau}^2 N_{\tau}}
        \sum_{a \in \urm{occ}}
        \ca{N}_{a \tau}
        \avr{\ve{l} \cdot \ve{\sigma}}, 
        \label{eq:second_SO} \\
      \avr{r^4}_{\urm{SO} \tau}
      & \simeq
        \frac{10}{M_{\tau}^2 N_{\tau}}
        \sum_{a \in \urm{occ}}
        \left[
        \frac{\kappa_{\tau}}{5}
        \avr{r^2}_{g_{a \tau}}
        +
        \frac{r_{\urm{M} \tau}^2 - r_{\urm{E} \tau}^2}{3}
        +
        \frac{\kappa_{\tau}}{2M_{\tau}^2}
        \right]
        \ca{N}_{a \tau}
        \avr{\ve{l} \cdot \ve{\sigma}},
        \label{eq:fourth_SO}
    \end{align}
  \end{subequations}
\end{widetext}
where
$ N_{\tau} = Z $ for proton ($ \tau = p $) or charge distribution
and
$ N_{\tau} = N $ for neutron ($ \tau = n $) distribution,
$ \kappa_{\tau} $ is
the anomalous magnetic moment of the nucleon $ \tau $~\cite{
  Zyla2020Prog.Theor.Exp.Phys.2020_083C01},
and $ r_{\urm{M} \tau}^2 $ is the second moment of magnetic distribution of the nucleon $ \tau $.
The index $ a = \left( n, \kappa, m \right) $ is the set of the quantum numbers of a single-particle orbital,
whose occupation number is $ \ca{N}_{a \tau} $, 
and $ \avr{r^2}_{g_{a \tau}} $ is the second moment of the radial part of the upper component of single-particle Dirac spinor $ g_{a \tau} \left( r \right) $,
which is approximately identical to the radial part of a single-particle orbital in the nonrelativistic scheme.
Detailed derivations are shown in Appendices~\ref{sec:appen_detail} and \ref{sec:appen_charge},
and see also Ref.~\cite{
  Horowitz2012Phys.Rev.C86_045503}
for Eq.~\eqref{eq:second_SO}.
\par
One can simply assume that $ \avr{r^2}_{g_{a \tau}} \simeq \avr{r^2}_{\tau} $,
which is probably a good approximation except in weakly bound systems, and estimate the spin-orbit contribution
$ \avr{r^2}_{\urm{SO} \tau} $ and $ \avr{r^4}_{\urm{SO} \tau} $
based on the naive shell-model occupancies.
The
approximation brings us to  
\begin{widetext}
  \begin{equation}
    \label{eq:fourth_SO_est}
    \avr{r^4}_{\urm{SO} \tau}
    \simeq 
    \frac{10}{M_{\tau}^2 N_{\tau}}
    \left[
      \frac{\kappa_{\tau}}{5}
      \avr{r^2}_{\tau}
      +
      \frac{r_{\urm{M} \tau}^2 - r_{\urm{E} \tau}^2}{3}
      +
      \frac{\kappa_{\tau}}{2M_{\tau}^2}
    \right] 
    \sum_{a \in \urm{occ}}
    \ca{N}_{a \tau}
    \avr{\ve{l} \cdot \ve{\sigma}}
  \end{equation}
\end{widetext}
and we will discuss below the role of the occupancies $ \ca{N}_{a \tau} $
(cf.~Sec.~\ref{sec:benchmark}).
\par
In this paper, we use the electric form factors of protons and neutrons
$ \tilde{G}_{\urm{E} \tau} $ proposed in Ref.~\cite{
  Alberico2009Phys.Rev.C79_065204}
in which values of $ r_{\urm{E} \tau}^2 $, $ r_{\urm{E} \tau}^4 $, and $ r_{\urm{M} \tau}^2 $ are
\begin{subequations}
  \begin{align}
    r_{\urm{E} \tau}^2
    & =
      \begin{cases}
        0.75036 \, \mathrm{fm}^2
        & \text{(Proton)}, \\
        -0.11146 \, \mathrm{fm}^2
        & \text{(Neutron)}, 
      \end{cases} \\
    r_{\urm{E} \tau}^4
    & =
      \begin{cases}
        1.6228 \, \mathrm{fm}^4
        & \text{(Proton)}, \\
        -0.33398 \, \mathrm{fm}^4
        & \text{(Neutron)}, 
      \end{cases} \\
    r_{\urm{M} \tau}^2
    & =
      \begin{cases}
        0.74439 \, \mathrm{fm}^2
        & \text{(Proton)}, \\
        0.86381 \, \mathrm{fm}^2
        & \text{(Neutron)}, 
      \end{cases}
  \end{align}
\end{subequations}
respectively.
Note that these values of $ r_{\urm{E} \tau}^2 $ and $ r_{\urm{M} \tau}^2 $ are accurate enough for our purpose~\cite{
  Zyla2020Prog.Theor.Exp.Phys.2020_083C01}
and
some form factors available in the literature~\cite{
  Friedrich2003Eur.Phys.J.A17_607}
give the opposite sign for $ r_{\urm{E} n}^4 $,
but this difference does not change the discussion presented in this paper.
% PDG
% rEp2 =  0.7071
% rEn2 = -0.1161
% 
\section{Isotope-shift Method}
\label{sec:diff}
\par
We then consider whether $ \avr{r^2}_n $ can be extracted from experimental data of $ \avr{r^2}_{\urm{ch}} $ and $ \avr{r^4}_{\urm{ch}} $
provided by the electron scattering experiments or isotope shift
by using Eqs.~\eqref{eq:second} and \eqref{eq:fourth}.
Although $ \avr{r^2}_{\urm{SO} \tau} $ and $ \avr{r^4}_{\urm{SO} \tau} $ are derived in Eqs.~\eqref{eq:second_SO} and \eqref{eq:fourth_SO_est},
several approximations have been introduced to derive them, as shown in Appendix~\ref{sec:appen_charge}.
To reduce uncertainties introduced by such approximations and by the nucleon form factors,
we will consider two isotopes with the neutron numbers $ N - 2 $ and $ N $,
instead of only one nucleus.
Since $ \avr{r^2}_{\urm{SO} \tau} $ and $ \avr{r^4}_{\urm{SO} \tau} $
are written as the sum of $ \avr{\ve{l} \cdot \ve{\sigma}} $ over all the single-particle orbitals,
$ \avr{r^n}_{\urm{SO} \tau} $ for two isotopes in the same neutron shell are almost the same.
Hence, a large cancellation of such uncertainty can be expected.
\par
Comparing Eq.~\eqref{eq:fourth} for two isotopes with their neutron numbers $ N - 2 $ and $ N $,
the master equation,
\begin{widetext}
  \begin{align}
    & \avr{r^2}_n^{\left( N - 2 \right)}
      \notag \\
    = & \,
        \frac{3}{10}
        \frac{Z}{r_{\urm{E} n}^2}
        \left[
        \frac{\avr{r^4}_{\urm{ch}}^{\left( N \right)} - \avr{r^4}_{\urm{ch}}^{\left( N - 2 \right)}}{2}
        -
        \frac{\avr{r^4}_p^{\left( N \right)} - \avr{r^4}_p^{\left( N - 2 \right)}}{2}
        -
        \frac{
        \avr{r^4}_{\urm{SO} p}^{\left( N \right)}
        -
        \avr{r^4}_{\urm{SO} p}^{\left( N - 2 \right)}}{2}
        \right]
        -
        Z
        \frac{r_{\urm{E} p}^2}{r_{\urm{E} n}^2}
        \frac{\avr{r^2}_{\urm{ch}}^{\left( N \right)} - \avr{r^2}_{\urm{ch}}^{\left( N - 2 \right)}}{2}
        \notag \\
    & \,
      -
      N
      \frac{\avr{r^2}_n^{\left( N \right)} - \avr{r^2}_n^{\left( N - 2 \right)}}{2}
      -
      \frac{3}{10}
      \frac{N}{r_{\urm{E} n}^2}
      \frac{
      \avr{r^4}_{\urm{SO} n}^{\left( N \right)}
      -
      \avr{r^4}_{\urm{SO} n}^{\left( N - 2 \right)}}{2}
      -
      \frac{3}{10}
      \frac{1}{r_{\urm{E} n}^2}
      \avr{r^4}_{\urm{SO} n}^{\left( N - 2 \right)}
      \notag \\
    & \, 
      +
      r_{\urm{E} p}^2
      +
      \frac{r_{\urm{E} p}^2}{r_{\urm{E} n}^2}
      \frac{
      N
      \avr{r^2}_{\urm{SO} n}^{\left( N \right)} 
      -
      \left( N - 2 \right)
      \avr{r^2}_{\urm{SO} n}^{\left( N-2 \right)}}{2}
      -
      \frac{3}{10}
      \frac{r_{\urm{E} n}^4}{r_{\urm{E} n}^2}
      \label{eq:master} 
  \end{align}
\end{widetext}
is derived,
where the superscripts $ \left( N - 2 \right) $ and $ \left( N \right) $ describe the quantities for the nuclei with the neutron numbers $ N - 2 $ and $ N $, respectively.
If one assumes the integer occupation with the standard shell structure,
$ \frac{N \avr{r^2}_{\urm{SO} n}^{\left( N \right)} 
  -
  \left( N - 2 \right) \avr{r^2}_{\urm{SO} n}^{\left( N-2 \right)}}{2} $ 
can be further simplified as
$ \frac{\kappa_n}{M_{\tau}^2}
\avr{\ve{l} \cdot \ve{\sigma}}_{n \urm{last}} $, 
where $ \avr{\ve{l} \cdot \ve{\sigma}}_{n \urm{last}} $ is $ \avr{\ve{l} \cdot \ve{\sigma}} $ for the orbital that the last neutron occupies.
On the right-hand side of Eq.~\eqref{eq:master},
$ \avr{r^4}_{\urm{ch}} $ is given by the experimental data,
and $ \avr{r^2}_p $ is given by Eq.~\eqref{eq:second} and the experimental value of $ \avr{r^2}_{\urm{ch}} $.
Meanwhile, the way to derive $ \avr{r^4}_p $, which is not known, will be discussed later.
The remaining term
$ N \frac{\avr{r^2}_n^{\left( N \right)} - \avr{r^2}_n^{\left( N - 2 \right)}}{2} $
will be shown to be small and almost model independent,
thus, we can adopt theoretically predicted values for this factor which will be referred to as the ``neutron slope term,''
as we explain in the following.
Since the spin-orbit contribution to the second moment $ \avr{r^2}_{\urm{SO} \tau} $ can be estimated 
and that to the fourth moment, $ \avr{r^4}_{\urm{SO} \tau} $, is much smaller than the other contributions as will be shown in Table~\ref{tab:benchmark},
we assume it is known.
\par
\begin{figure}[tb]
  \centering
  \includegraphics[width=1.0\linewidth]{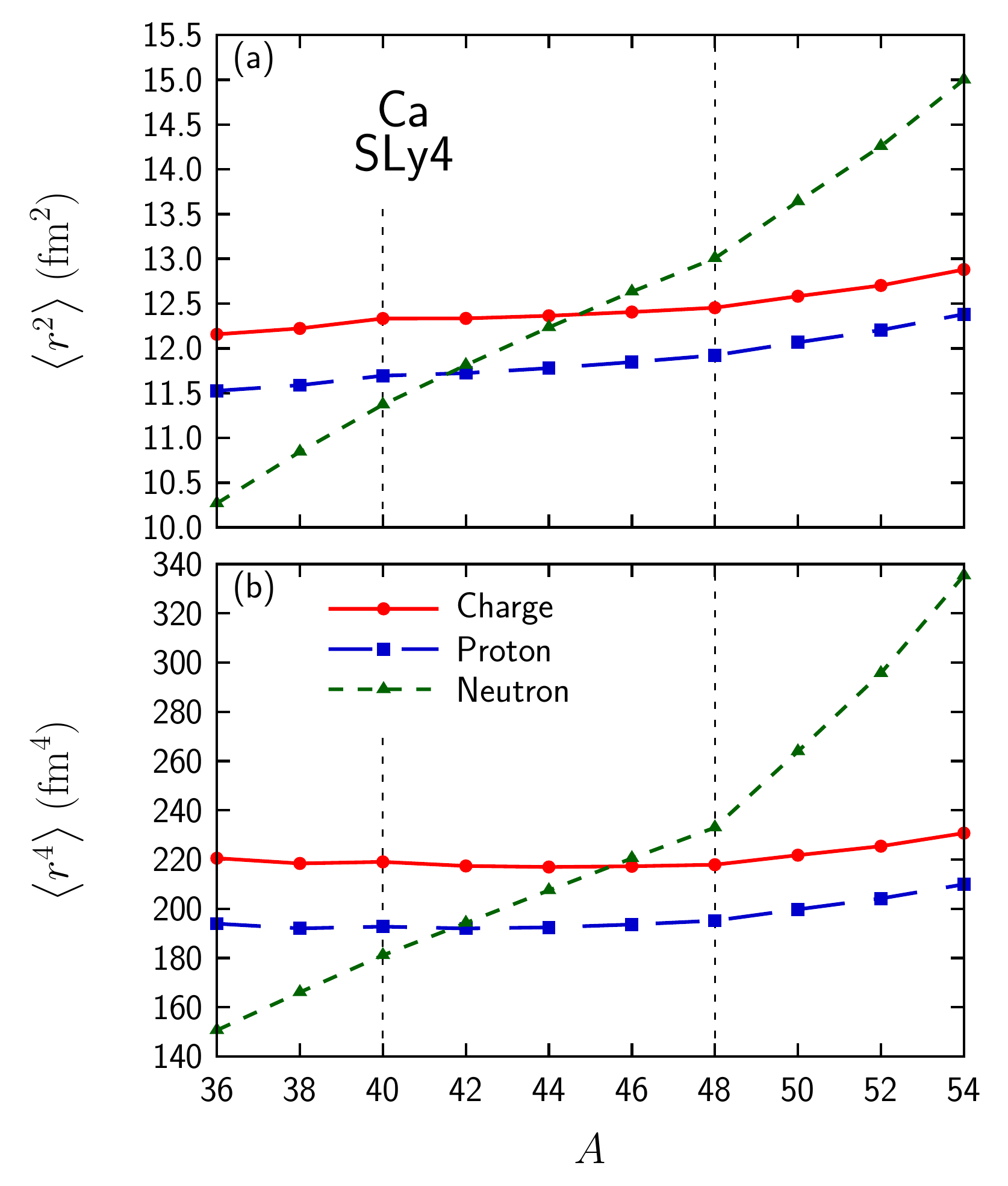}
  \caption{The second $ \avr{r^2} $ and fourth $ \avr{r^4} $ moments of
    the proton, neutron, and charge density distributions
    of $ \mathrm{Ca} $ isotopes
    as functions of mass number $ A $.
    They are shown with blue long-dashed, green dashed, and red solid lines, respectively.
    As an example, the SLy4 functional~\cite{
      Chabanat1998Nucl.Phys.A635_231} is used.}
  \label{fig:Ca_isotope}
\end{figure}
\begin{figure}[tb]
  \centering
  \includegraphics[width=1.0\linewidth]{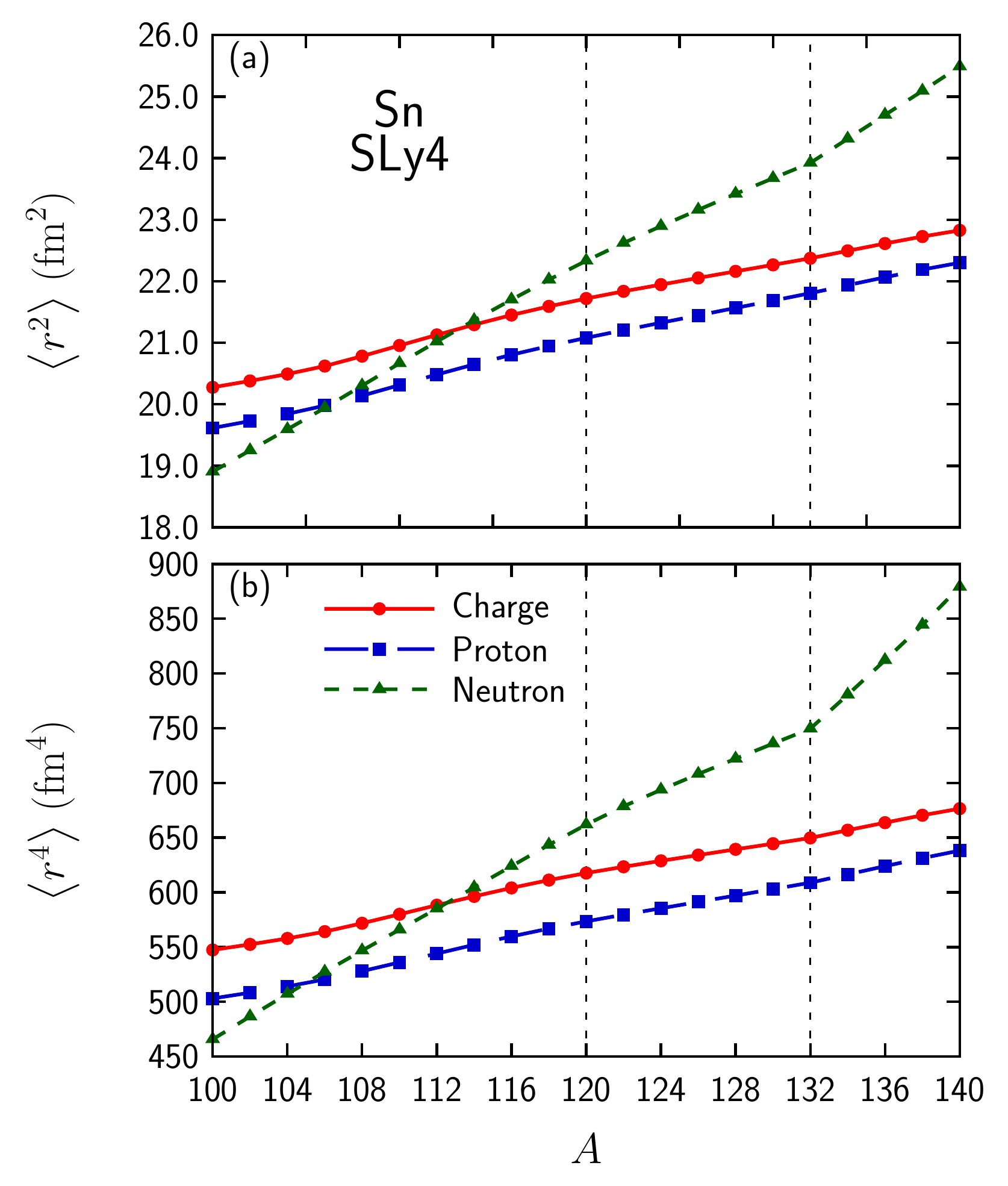}
  \caption{Same as Fig.~\ref{fig:Ca_isotope} but for $ \mathrm{Sn} $ isotopes.}
  \label{fig:Sn_isotope}
\end{figure}

The slopes of the second moments in the same neutron (sub)shell,
e.g.,~$ A \in \left[ 40, 48 \right] $ for $ \mathrm{Ca} $ isotopes
or $ A \in \left[ 100, 120 \right] $ for $ \mathrm{Sn} $ isotopes,
are almost constant.
In other words,
$ \avr{r^2}_n^{\left( N \right)} - \avr{r^2}_n^{\left( N - 2 \right)} $
or $ N \left( \avr{r^2}_n^{\left( N \right)} - \avr{r^2}_n^{\left( N - 2 \right)} \right) $
is almost constant as seen in Figs.~\ref{fig:Ca_isotope} and \ref{fig:Sn_isotope}.
By studying the predictions of several models 
for the neutron slope term as well,
we have found that it is almost model independent.
Hence,
in the benchmark calculation in the next section,
we will use the averaged value of
$ N \left( \avr{r^2}_n^{\left( N \right)} - \avr{r^2}_n^{\left( N - 2 \right)} \right) $
among the values for the same neutron (sub)shell
($ N = 22 $, $ 24 $, $ 26 $, and $ 28 $ for $ \mathrm{Ca} $ isotopes
and
$ N = 52 $, $ 54 $, $ 56 $, \ldots, $ 70 $ for $ \mathrm{Sn} $ isotopes)
calculated with the selected energy density functionals.
As we will show, our mild assumptions on the neutron slope and spin-orbit contributions
will not affect our conclusions.  
\section{Benchmark calculation}
\label{sec:benchmark}
\par
As a benchmark calculation, we test whether the neutron radius calculated theoretically can be reproduced in this novel method.
During the benchmark calculation, the theoretical values of $ \avr{r^2}_{\urm{ch}} $ and $ \avr{r^4}_{\urm{ch}} $ are used,
and we will see how accurately $ \avr{r^2}_n^{\left( N - 2 \right)} $ can be calculated from Eq.~\eqref{eq:master},
or how large the neutron slope term and the spin-orbit term or the pairing introduce an uncertainty.
\par
The Skyrme Hartree-Fock-Bogoliubov calculation~\cite{
  Vautherin1972Phys.Rev.C5_626,
  Dobaczewski1984Nucl.Phys.A422_103}
is performed under the assumption of the axial symmetry
using the code \textsc{hfbtho}~\cite{
  NavarroPerez2017Comput.Phys.Commun.220_363}.
The calculations are performed using a basis of the spherical harmonic oscillator
in which $ 24 $ major shells are taken into account
and
whose oscillator frequency $ \omega_0 $ satisfies
$ \hbar \omega_0 = 1.2 \times 41 A^{-1/3} \, \mathrm{MeV} $.
As for the proton-proton and neutron-neutron pairing force, a volume-type pairing force~\cite{
  Dobaczewski1984Nucl.Phys.A422_103}
\begin{equation}
  \label{eq:pairing}
  V_{\urm{pair} \tau} \left( \ve{r}, \ve{r}' \right)
  =
  - V_{0 \tau}
  \delta \left( \ve{r} - \ve{r}' \right)
\end{equation}
is used
where the pairing strength $ V_{0p} = V_{0n} $ is determined to reproduce the pairing gap of $ \nuc{Sn}{120}{} $ as $ 1.4 \, \mathrm{MeV} $
with the cutoff energy in quasiparticle space $ 60 \, \mathrm{MeV} $.
The
SLy4~\cite{
  Chabanat1998Nucl.Phys.A635_231},
SLy5~\cite{
  Chabanat1998Nucl.Phys.A635_231},
SkM*~\cite{
  Bartel1982Nucl.Phys.A386_79},
SAMi~\cite{
  Roca-Maza2012Phys.Rev.C86_031306},
HFB9~\cite{
  Goriely2005Nucl.Phys.A750_425},
UNEDF0~\cite{
  Kortelainen2010Phys.Rev.C82_024313}, 
UNEDF1~\cite{
  Kortelainen2012Phys.Rev.C85_024304},
and
UNEDF2~\cite{
  Kortelainen2014Phys.Rev.C89_054314}
EDFs are used as examples,
whose pairing strengths $ V_{0 \tau} $ are
$ 194.2 $, $ 188.2 $, $ 156.2 $, $ 213.7 $,
$ 166.4 $, $ 127.6 $, $ 138.4 $, and $ 150.0 \, \mathrm{MeV} \, \mathrm{fm}^3 $,
respectively.
\begin{table*}[tb]
  \centering
  \caption{Benchmark calculation results of
    $ \avr{r^2}_{\tau} $, $ \avr{r^4}_{\tau} $,
    $ \avr{r^2}_{\urm{SO} \tau} $, $ \avr{r^4}_{\urm{SO} \tau} $,
    $ \avr{r^2}_{\urm{ch}} $, and $ \avr{r^4}_{\urm{ch}} $.
    The SLy4 energy density functional \cite{
      Chabanat1998Nucl.Phys.A635_231} is used to calculate
    $ \avr{r^2}_{\tau} $ and $ \avr{r^4}_{\tau} $.
    The spin-orbit contributions,
    $ \avr{r^2}_{\urm{SO} \tau} $ and $ \avr{r^4}_{\urm{SO} \tau} $,
    are calculated by using Eqs.~\eqref{eq:second_SO} and \eqref{eq:fourth_SO_est}.
    See the text for detail.}
  \label{tab:benchmark}
  \begin{ruledtabular}
    \begin{tabular}{ldddddddddd}
      \multicolumn{1}{c}{\multirow{2}{*}{Isotope}} & \multicolumn{5}{c}{Second moments ($ \mathrm{fm}^2 $)} & \multicolumn{5}{c}{Fourth moments ($ \mathrm{fm}^4 $)} \\
      \cline{2-6} \cline{7-11}
                                                   & \multicolumn{1}{c}{$ \avr{r^2}_p $} & \multicolumn{1}{c}{$ \avr{r^2}_n $} & \multicolumn{1}{c}{$ \avr{r^2}_{\urm{SO} p} $} & \multicolumn{1}{c}{$ \frac{N}{Z} \avr{r^2}_{\urm{SO} n} $} & \multicolumn{1}{c}{$ \avr{r^2}_{\urm{ch}} $} & \multicolumn{1}{c}{$ \avr{r^4}_p $} & \multicolumn{1}{c}{$ \avr{r^4}_n $} & \multicolumn{1}{c}{$ \avr{r^4}_{\urm{SO} p} $} & \multicolumn{1}{c}{$ \frac{N}{Z} \avr{r^4}_{\urm{SO} n} $} & \multicolumn{1}{c}{$ \avr{r^4}_{\urm{ch}} $} \\
      \hline
      % $ \nuc{Ca}{44}{} $  & 11.778337 & 12.233967 & 0.0000000 & -0.0506285 & 12.344317 & 192.444618 & 207.485688 & 0.0 & -1.16391 & 216.508365 \\
      % $ \nuc{Ca}{46}{} $  & 11.847033 & 12.632462 & 0.0000000 & -0.0759427 & 12.376553 & 193.613271 & 220.471450 & 0.0 & -1.80638 & 216.525965 \\
      % \hline
      % $ \nuc{Sn}{110}{} $ & 20.312464 & 20.665109 & 0.0634379 & -0.00675046 & 20.985760 & 535.964654 & 565.926186 & 2.67654 & -0.269016 & 581.186456 \\
      % $ \nuc{Sn}{112}{} $ & 20.483428 & 21.019612 & 0.0634379 & -0.0135009  & 21.145515 & 544.080290 & 585.110959 & 2.69744 & -0.547603 & 588.988228 \\
      % 
      $ \nuc{Ca}{44}{} $  & 11.778 & 12.234 & 0.000 & -0.051 & 12.344 & 192.445 & 207.486 & 0.000 & -1.162 & 216.510 \\
      $ \nuc{Ca}{46}{} $  & 11.847 & 12.632 & 0.000 & -0.075 & 12.377 & 193.613 & 220.471 & 0.000 & -1.792 & 216.540 \\
      \hline
      $ \nuc{Sn}{110}{} $ & 20.312 & 20.665 & 0.063 & -0.058 & 20.934 & 535.965 & 565.926 & 2.677 & -2.320 & 579.136 \\
      $ \nuc{Sn}{112}{} $ & 20.483 & 21.020 & 0.063 & -0.049 & 21.110 & 544.080 & 585.111 & 2.697 & -1.987 & 587.549 \\
    \end{tabular}
  \end{ruledtabular}
\end{table*}
\par
In this benchmark calculation,
first, $ \avr{r^2}_{\tau} $ and $ \avr{r^4}_{\tau} $
are calculated by the \textsc{hfbtho} code,
and $ \avr{r^2}_{\urm{ch}} $ and $ \avr{r^4}_{\urm{ch}} $
are evaluated by using Eqs.~\eqref{eq:second}, \eqref{eq:fourth}, \eqref{eq:second_SO}, and \eqref{eq:fourth_SO_est}.
Then, $ \avr{r^2}_{\urm{ch}} $ and $ \avr{r^4}_{\urm{ch}} $ are assumed to be known,
and we test how accurately $ \avr{r^2}_{\tau} $ and $ \avr{r^4}_{\tau} $ can be evaluated.
Sources of uncertainty discussed in this paper are
estimation of $ \avr{r^4}_p $ and the neutron slope term, 
and 
if no uncertainty is introduced, 
evaluated $ \avr{r^2}_n $ should be consistent to that calculated by the \textsc{hfbtho} code.
The contributions of $ \avr{r^2}_{\urm{SO} \tau} $ and $ \avr{r^4}_{\urm{SO} \tau} $ 
are estimated as previously explained and will not play a prominent role in the determination of
$ \avr{r^2}_n $ when compared with other sources of uncertainties,
and thus they are assumed to be known.
As examples,
the $ \mathrm{Ca} $ isotope with $ N = 26 $
and
the $ \mathrm{Sn} $ isotope with $ N = 62 $
are chosen,
i.e.,
($ \nuc{Ca}{44}{}  $, $ \nuc{Ca}{46}{}  $) and
($ \nuc{Sn}{110}{} $, $ \nuc{Sn}{112}{} $) pairs are used for Eq.~\eqref{eq:master}.
The values calculated with the SLy4 EDF are used.
Calculation results of $ \avr{r^2}_{\tau} $ and $ \avr{r^4}_{\tau} $ are shown in Table~\ref{tab:benchmark}.
Using these results, accordingly, we calculate
$ \avr{r^2}_{\urm{SO} \tau} $, $ \avr{r^4}_{\urm{SO} \tau} $,
$ \avr{r^2}_{\urm{ch}} $, and $ \avr{r^4}_{\urm{ch}} $ as shown in Table~\ref{tab:benchmark},
where $ \sum_a \ca{N}_{a \tau} \avr{\ve{l} \cdot \ve{\sigma}} $ are derived from the results of Hartree-Fock-Bogoliubov calculation
as shown in Table~\ref{tab:occupation}.
For comparison, Table~\ref{tab:occupation} shows $ \sum_a \ca{N}_{a \tau} \avr{\ve{l} \cdot \ve{\sigma}} $
calculated by using the Hartree-Fock calculation, i.e., integer $ \ca{N}_{a \tau} $.
On the one hand, since $ \mathrm{Ca} $ and $ \mathrm{Sn} $ are proton magic nuclei,
the proton pairing does not change $ \sum_a \ca{N}_{a \tau} \avr{\ve{l} \cdot \ve{\sigma}} $ for protons.
On the other hand, one can find that the neutron pairing affects $ \sum_a \ca{N}_{a \tau} \avr{\ve{l} \cdot \ve{\sigma}} $ for neutrons, at most, approximately $ 30 \, \% $,
and that the resulting impact on $ \avr{r^2}_{\urm{ch}} $ and $ \avr{r^4}_{\urm{ch}} $ is
eventually less than $ 0.5 \, \% $.
Thus, as discussed later, uncertainties associated with the spin-orbit contribution due to the pairing are negligible,
and hereinafter, this will not be considered.
\begin{table}[tb]
  \centering
  \caption{Spin-orbit expectation values $ \sum_a \ca{N}_{a \tau} \avr{\ve{l} \cdot \ve{\sigma}} $ calculated by using the Hartree-Fock-Bogoliubov method.
    For comparison, those calculated by using the Hartree-Fock method (integer $ \ca{N}_{a \tau} $) are also shown.}
  \label{tab:occupation}
  \begin{ruledtabular}
    \begin{tabular}{ldddd}
      \multicolumn{1}{c}{\multirow{2}{*}{Nuclei}} & \multicolumn{2}{c}{Proton $ \sum_a \ca{N}_{a \tau} \avr{\ve{l} \cdot \ve{\sigma}} $} & \multicolumn{2}{c}{Neutron $ \sum_a \ca{N}_{a \tau} \avr{\ve{l} \cdot \ve{\sigma}} $} \\
      \cline{2-3} \cline{4-5}
                                                  & \multicolumn{1}{c}{HF} & \multicolumn{1}{c}{HFB} & \multicolumn{1}{c}{HF} & \multicolumn{1}{c}{HFB} \\
      \hline
      $ \nuc{Ca}{44}{} $  &   0 &   0.00 & +12 & +11.982 \\
      $ \nuc{Ca}{46}{} $  &   0 &   0.00 & +18 & +17.860 \\
      $ \nuc{Sn}{110}{} $ & +40 & +40.00 & +32 & +34.492 \\
      $ \nuc{Sn}{112}{} $ & +40 & +40.00 & +22 & +29.025 \\
    \end{tabular}
  \end{ruledtabular}
\end{table}
\par
Figures~\ref{fig:Ca_isotope} and \ref{fig:Sn_isotope}, respectively, show
$ \avr{r^2} $ and $ \avr{r^4} $ 
of $ \mathrm{Ca} $ and $ \mathrm{Sn} $ isotopes
calculated with the SLy4 EDF 
as functions of the mass number $ A $.
It should be noted that all the calculation results are eventually spherical,
although the axial deformation is allowed in the numerical calculations.
\subsection{Derivation of proton fourth moment}
\label{sec:R4p}
\par
Before going into the discussion on the neutron slope term, 
to derive $ \avr{r^2}_n $ from Eqs.~\eqref{eq:second} and \eqref{eq:fourth},
$ \avr{r^4}_p $ should be estimated in a certain way
since it cannot be determined from experimental data, in contrast to $ \avr{r^2}_p $.
In this paper, we adopt a way which 
was 
similar to that used in Ref.~\cite{
  Kurasawa2021Prog.Theor.Exp.Phys.2021_013D02}.
\par
\begin{figure*}[htb]
  \centering
  \begin{minipage}{0.49\linewidth}
    \centering
    \includegraphics[width=1.0\linewidth]{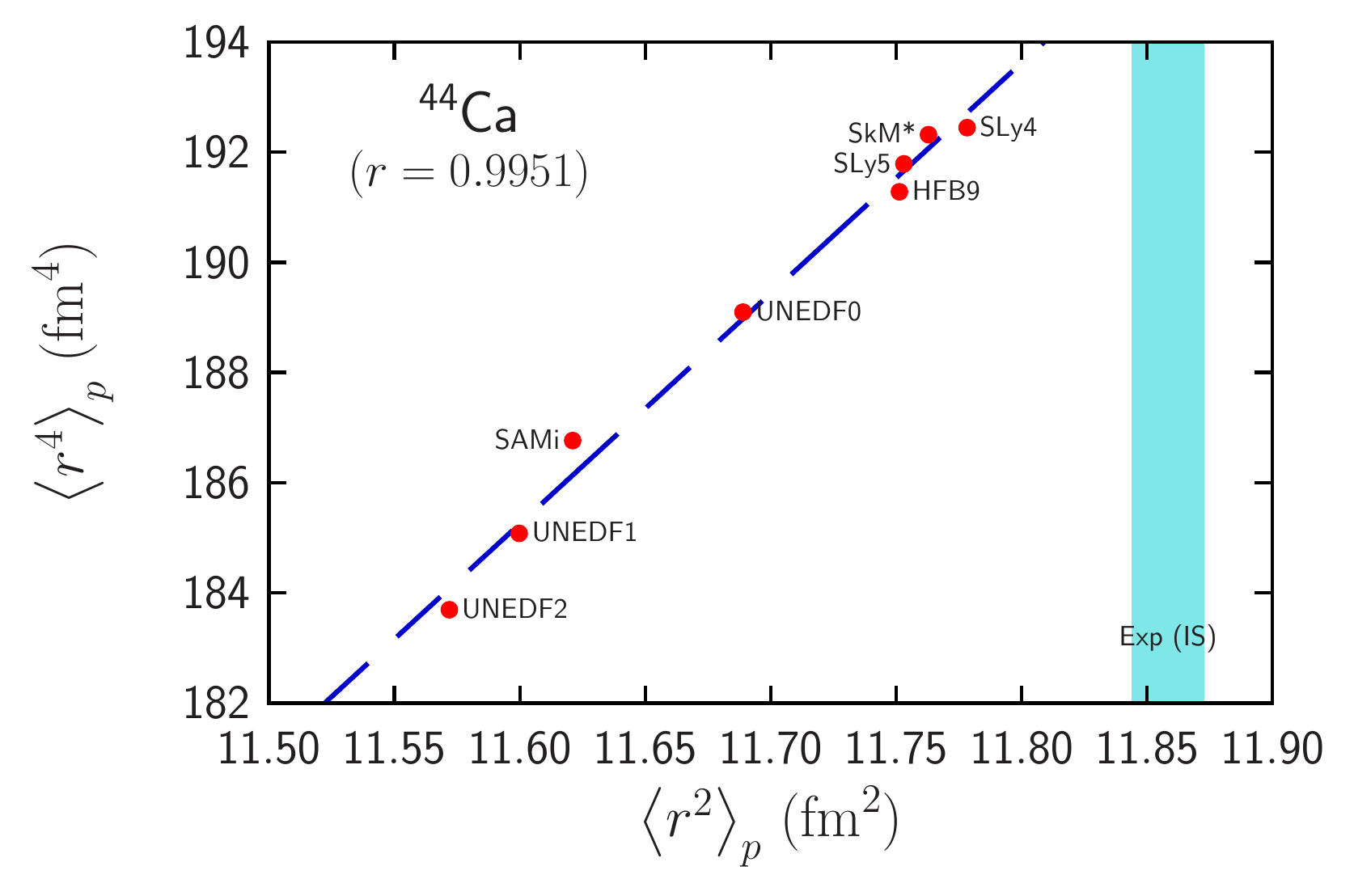}
  \end{minipage}
  \hfill
  \begin{minipage}{0.49\linewidth}
    \centering
    \includegraphics[width=1.0\linewidth]{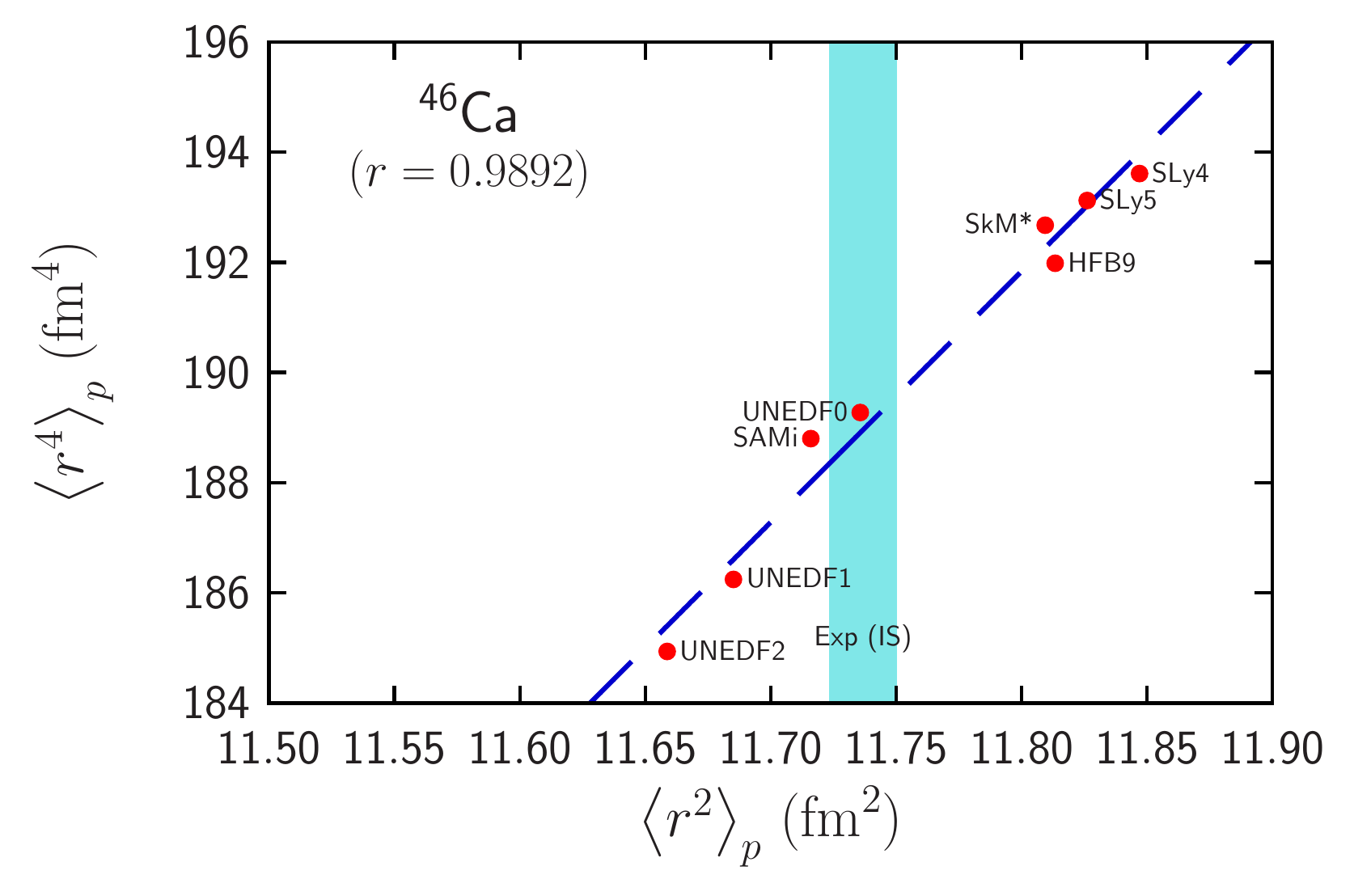}
  \end{minipage}
  \begin{minipage}{0.49\linewidth}
    \centering
    \includegraphics[width=1.0\linewidth]{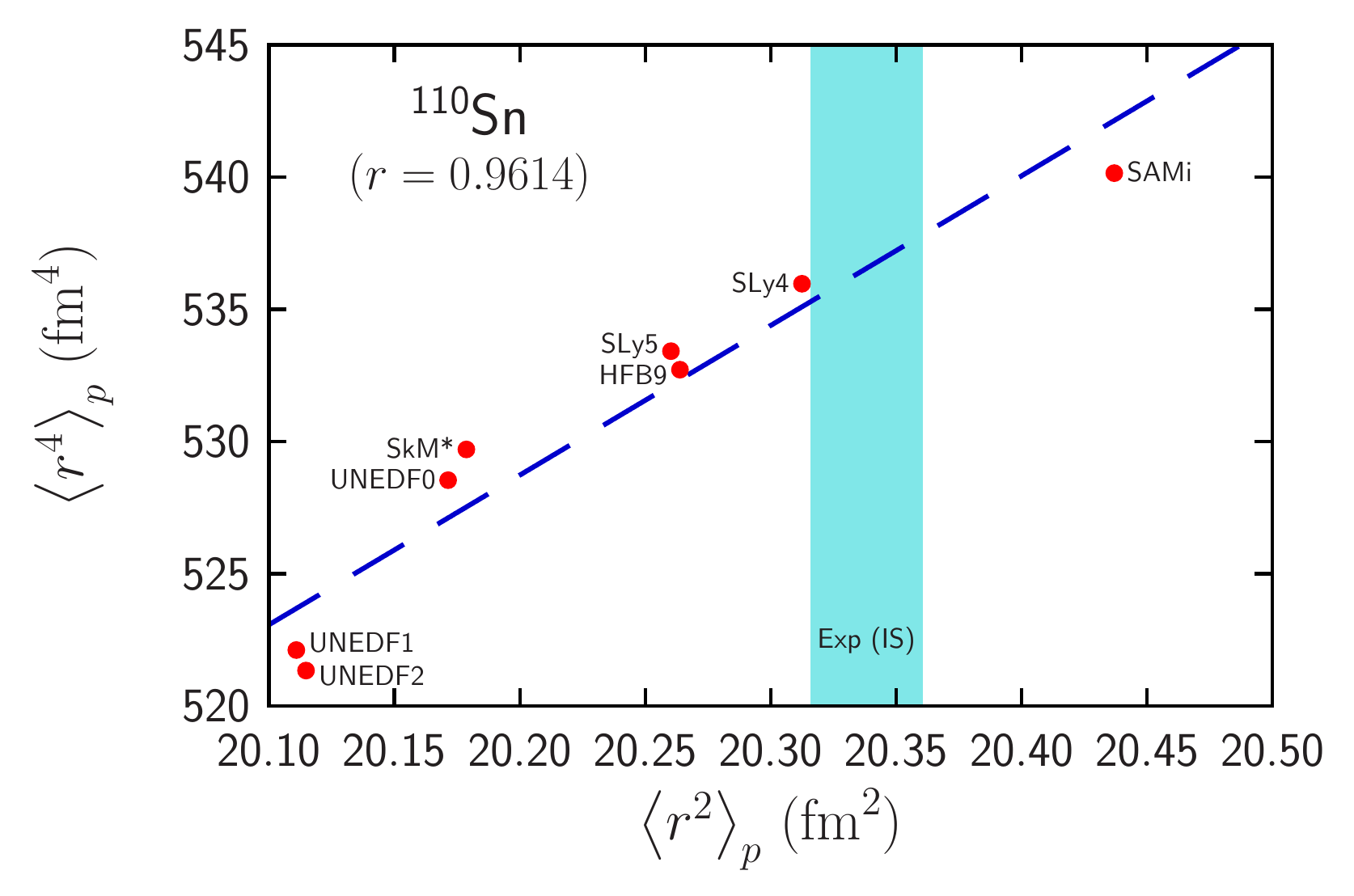}
  \end{minipage}
  \hfill
  \begin{minipage}{0.49\linewidth}
    \centering
    \includegraphics[width=1.0\linewidth]{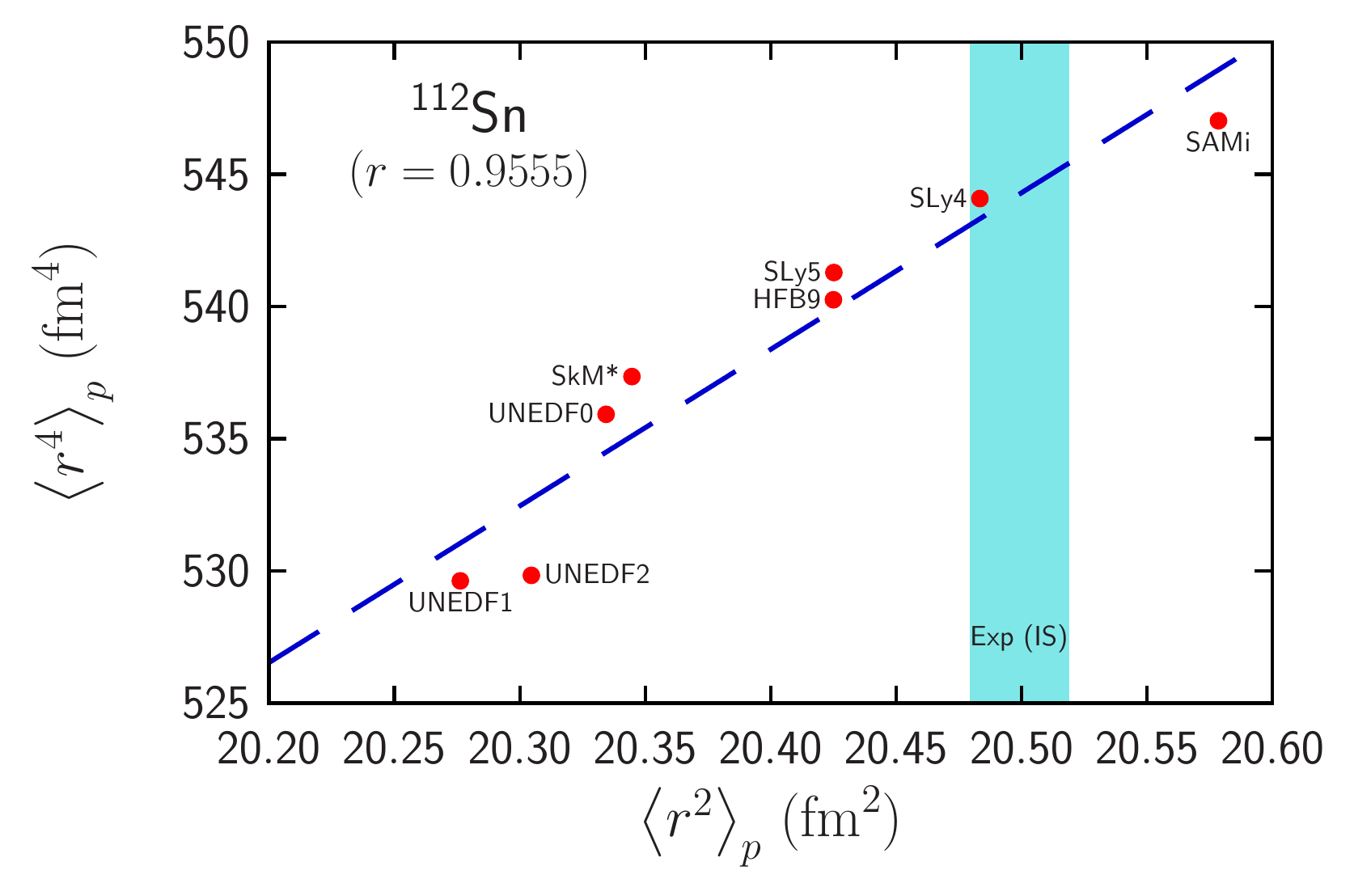}
  \end{minipage}
  \caption{Correlation of proton second and fourth moments, $ \avr{r^2}_p $ and $ \avr{r^4}_p $
    for $ \nuc{Ca}{44}{} $, $ \nuc{Ca}{46}{} $, $ \nuc{Sn}{110}{} $, and $ \nuc{Sn}{112}{} $.
    The proton second moments extracted from experimental charge radii measured by using isotope shift (IS) method~\cite{
      Angeli2013At.DataNucl.DataTables99_69}
    and Eq.~\eqref{eq:second} are also shown as filled bands.
    See the text for more details.}
  \label{fig:proton}
\end{figure*}

We estimate the correlation between $ \avr{r^2}_p $ and $ \avr{r^4}_p $ for $ \nuc{Ca}{44}{} $, $ \nuc{Ca}{46}{} $, $ \nuc{Sn}{110}{} $, and $ \nuc{Sn}{112}{} $ 
by using the theoretical results for the selected EDFs as 
\begin{subequations}
  \label{eq:R4p}
  \begin{align}
    \avr{r^4}_p^{\urm{Ca-44}}
    = & \, 
    % \left( 41.83764 \pm 1.703847 \right)
        \left( 41.838 \pm 1.704 \right)
        \avr{r^2}_p^{\urm{Ca-44}}
        \notag \\
      & \,
        -
        % \left( 300.0623 \pm 19.91999 \right) 
        \left( 300.062 \pm 19.920 \right) 
        \qquad
        \text{($ r = 0.9951 $)}, 
        \label{eq:R4p_Ca44} \\
    \avr{r^4}_p^{\urm{Ca-46}}
    = & \, 
    % \left( 45.55866 \pm 2.759620 \right)
        \left( 45.559 \pm 2.760 \right)
        \avr{r^2}_p^{\urm{Ca-46}}
        \notag \\
      & \, 
        -
        % \left( 345.7548 \pm 32.45769 \right) 
        \left( 345.755 \pm 32.458 \right) 
        \qquad
        \text{($ r = 0.9892 $)}, 
        \label{eq:R4p_Ca46} \\
    \avr{r^4}_p^{\urm{Sn-110}}
    = & \, 
    % \left( 56.58531 \pm 6.607341 \right)
        \left( 56.585 \pm 6.607 \right)
        \avr{r^2}_p^{\urm{Sn-110}}
        \notag \\
      & \,
        -
        % \left( 614.2959 \pm 133.6760 \right) 
        \left( 614.296 \pm 133.676 \right) 
        \qquad
        \text{($ r = 0.9614 $)}, 
        \label{eq:R4p_Sn110} \\
    \avr{r^4}_p^{\urm{Sn-112}}
    = & \,
    % \left( 59.22340 \pm 7.461120 \right)
        \left( 59.223 \pm 7.461 \right)
        \avr{r^2}_p^{\urm{Sn-112}}
        \notag \\
      & \, 
        -
        % \left( 669.7812 \pm 152.1825 \right)
        \left( 669.781 \pm 152.183 \right)
        \qquad
        \text{($ r = 0.9555 $)},
        \label{eq:R4p_Sn112} 
  \end{align}
\end{subequations}
respectively, as shown in Fig.~\ref{fig:proton}.
The proton second moments extracted from experimental charge radii~\cite{
  Angeli2013At.DataNucl.DataTables99_69},
experimental nucleon second moments
($ r_{\urm{E} p} = 0.8409 \pm 0.0004 \, \mathrm{fm} $
and $ r_{\urm{E} n}^2 = -0.1161 \pm 0.0022 \, \mathrm{fm}^2 $)~\cite{
  Zyla2020Prog.Theor.Exp.Phys.2020_083C01}, 
and Eq.~\eqref{eq:second} are also shown as filled bands
where Hartree-Fock (integer) occupations are used for $ \ca{N}_{a \tau} $.
Given that the estimated values of $ \avr{r^2}_p $ shown in Table~\ref{tab:benchmark}
are 
$ \avr{r^2}_p^{\urm{Ca-44}} = 11.778 \, \mathrm{fm}^2 $,
$ \avr{r^2}_p^{\urm{Ca-46}}  = 11.847 \, \mathrm{fm}^2 $,
$ \avr{r^2}_p^{\urm{Sn-110}} = 20.312 \, \mathrm{fm}^2 $, and
$ \avr{r^2}_p^{\urm{Sn-112}} = 20.483 \, \mathrm{fm}^2 $,
we infer that the estimated values of $ \avr{r^4}_p $
for $ \nuc{Ca}{44}{} $, $ \nuc{Ca}{46}{} $, $ \nuc{Sn}{110}{} $, and $ \nuc{Sn}{112}{} $ are
% $ 192.715557 \pm  1.992676 \, \mathrm{fm}^4 $, 
% $ 193.980227 \pm  4.799726 \, \mathrm{fm}^4 $, 
% $ 535.091105 \pm 37.197751 \, \mathrm{fm}^4 $, 
% $ 543.317104 \pm 45.482599 \, \mathrm{fm}^4 $,
$ 192.716 \pm  1.993 \, \mathrm{fm}^4 $, 
$ 193.980 \pm  4.800 \, \mathrm{fm}^4 $, 
$ 535.091 \pm 37.198 \, \mathrm{fm}^4 $, and
$ 543.317 \pm 45.483 \, \mathrm{fm}^4 $,
respectively.
These uncertainties are theoretical ones coming from the linear fits.
These uncertainties range 
from $ 1.03 \, \% $ to $ 8.37 \, \% $ 
and these deviations from the benchmark values are
around $ 0.15 \, \% $.
These good correlations ($ r \gtrsim 0.95 $) are due to the fact that the density profiles calculated with different EDFs share similar properties due to shell and orbital structures.
\par
Despite the good correlations between $ \avr{r^2}_p $ and $ \avr{r^4}_p $,
the final values of $ \avr{r^2}_n $ calculated by using Eq.~\eqref{eq:fourth},
or rearranged equation,
\begin{widetext}
  \begin{align}
    \avr{r^2}_n
    = & \, 
        \frac{3 Z}{10 N r_{\urm{E} n}^2}
        \left[
        \avr{r^4}_{\urm{ch}}
        -
        \avr{r^4}_p
        -
        \frac{10}{3}
        r_{\urm{E} p}^2
        \left\{
        \avr{r^2}_{\urm{ch}}
        -
        \left(
        r_{\urm{E} p}^2
        +
        \frac{N}{Z}
        r_{\urm{E} n}^2
        \right)
        -
        \avr{r^2}_{\urm{SO} p}
        -
        \frac{N}{Z}
        \avr{r^2}_{\urm{SO} n}
        \right\}
        \right.
        \notag \\
      & \,
        \left.
        -
        \left(
        r_{\urm{E} p}^4
        +
        \frac{N}{Z}
        r_{\urm{E} n}^4
        \right)
        -
        \avr{r^4}_{\urm{SO} p}
        -
        \frac{N}{Z}
        \avr{r^4}_{\urm{SO} n}
        \right],
        \label{eq:fourth_rea}
  \end{align}
\end{widetext}
are
$ \avr{r^2}_n^{\urm{Ca-44}}  = 12.842 \pm  4.470 \, \mathrm{fm}^2 $ and 
$ \avr{r^2}_n^{\urm{Sn-110}} = 18.706 \pm 83.434 \, \mathrm{fm}^2 $,
and, consequently,
$ \sqrt{\avr{r^2}_n^{\urm{Ca-44}}}  = 3.584 \pm 0.624 \, \mathrm{fm} $ and 
$ \sqrt{\avr{r^2}_n^{\urm{Sn-110}}} = 4.325 \pm 9.645 \, \mathrm{fm} $,
respectively,
whereas the benchmarked values are
$ \sqrt{\avr{r^2}_n^{\urm{Ca-44}}}  = 3.498 \, \mathrm{fm} $ and 
$ \sqrt{\avr{r^2}_n^{\urm{Sn-110}}} = 4.546 \, \mathrm{fm} $.
The uncertainties range
from $ 17 \, \% $ to $ 220 \, \% $.
This means the uncertainty is too large to extract $ \Delta r_{np} $ or even $ \sqrt{\avr{r^2}_n} $.
The reason why the error of $ \avr{r^2}_n $ enhances is due to the coefficient of $ \avr{r^4}_p $,
that is, $ 3 Z / \left| 10 N r_{\urm{E} n}^2 \right| \gtrsim 2.69 \, \mathrm{fm}^{-2} $.
\par
In short, extracting $ \avr{r^2}_n $ from the charge second and fourth moments is not feasible,
unless the proton fourth moment $ \avr{r^4}_p $ can also be determined precisely either experimentally or theoretically.
\subsection{Neutron slope term and isotope-shift method}
\label{sec:isotope_shift}
\par
In the previous section, we see that extracting $ \avr{r^2}_n $ from $ \avr{r^2}_{\urm{ch}} $ and $ \avr{r^4}_{\urm{ch}} $ may not be feasible.
Nevertheless, in this section, the other method called the isotope-shift method,
which is introduced in Sec.~\ref{sec:diff}, will be further discussed
since once $ \avr{r^4}_p $ is determined precisely, the method helps us to reduce uncertainty.
Furthermore, in laser spectroscopic experiments,
the difference in $ \avr{r^2}_{\urm{ch}} $ between two isotopes is obtained
whereas absolute values are not.
Thus, this isotope-shift method is still important for discussion.
\par 
The neutron slope term for $ \mathrm{Ca} $ and $ \mathrm{Sn} $ isotopes 
are derived by the average of $ 32 $ and $ 80 $ results,
which are for
$ N = 22 $, $ 24 $, $ 26 $, and $ 28 $ ($ \mathrm{Ca} $ isotopes) or 
$ N = 52 $, $ 54 $, $ 56 $, \ldots, $ 70 $ ($ \mathrm{Sn} $ isotopes)
calculated with
the selected eight functionals,
i.e., SLy4, SLy5, SkM*, SAMi, HFB9, UNEDF0, UNEDF1, and UNEDF2 \footnote{
  Lipkin-Nogami prescription is not used, while UNEDF series are fitted with the prescription.}.
The calculated value of $ N \left( \avr{r^2}_n^{\left( N \right)} - \avr{r^2}_n^{\left( N - 2 \right)} \right) $ is 
\begin{align}
  & N
    \left(
    \avr{r^2}_n^{\left( N \right)} - \avr{r^2}_n^{\left( N - 2 \right)}
    \right)
    \notag \\
  & =
  % \begin{cases}
  %   10.72025 \pm 1.155412  \qquad \text{($ \sigma^2 = 1.334977  $)}
  %   & \text{for $ \mathrm{Ca} $}, \\
  %   20.61435 \pm 2.085618  \qquad \text{($ \sigma^2 = 4.349804  $)}
  %   & \text{for $ \mathrm{Sn} $}
  % \end{cases}
        \begin{cases}
          10.720 \pm 1.155 \, \mathrm{fm}^2
          & \text{for $ \mathrm{Ca} $ isotopes}, \\
          20.614 \pm 2.086 \, \mathrm{fm}^2
          & \text{for $ \mathrm{Sn} $ isotopes},
        \end{cases}
            \label{eq:slope_2}
\end{align}
respectively.
Substituting
the neutron slope term [Eq.~\eqref{eq:slope_2}]
and
$ \avr{r^4}_p $ calculated in Sec.~\ref{sec:R4p}
as well as $ \avr{r^4}_{\urm{ch}} $ and $ \avr{r^2}_p $ shown in Table~\ref{tab:benchmark},
into Eq.~\eqref{eq:master},
we get $ \avr{r^2}_n $ of $ \nuc{Ca}{44}{} $ and $ \nuc{Sn}{110}{} $ as
\begin{subequations}
  \label{eq:neutron_2}
  \begin{align}
    \avr{r^2}_n^{\urm{Ca-44}}
    & =
      14.639 \pm  139.880 
      \, \mathrm{fm}^2, \\
    \avr{r^2}_n^{\urm{Sn-110}}
    & =
      28.774 \pm 3953.686
      \, \mathrm{fm}^2,
  \end{align}
\end{subequations}
respectively,
where breakdown of these uncertainties are shown in Table~\ref{tab:uncertainty}.
Accordingly, the neutron radii are calculated as
\begin{subequations}
  \label{eq:neutron_4}
  \begin{align}
    \sqrt{\avr{r^2}_n^{\urm{Ca-44}}}
    & =
      3.826 \pm  18.280
      \, \mathrm{fm}, \\
    \sqrt{\avr{r^2}_n^{\urm{Sn-110}}}
    & =
      5.364 \pm 368.531
      \, \mathrm{fm}.
  \end{align}
\end{subequations}
Obviously, the uncertainties are too large to extract $ \avr{r^2}_n $ and $ \sqrt{\avr{r^2}_n} $.
Note that the errors that are shown here are simply standard deviations.
For more details, see Appendix~\ref{sec:appen_error}.
\par
Contribution to these standard deviations can be divided into two parts:
the part that originates from $ \avr{r^4}_p $ and that from the neutron slope term.
Other sources can be considered as negligible.
Contribution of the neutron slope term to the total standard deviation $ \sigma^2 $
is approximately
$ 2 \, \mathrm{ppm} $ 
or less.
If there were no uncertainties due to $ \avr{r^4}_p $,
the results would become much improved as
\begin{subequations}
  \label{eq:neutron_6}
  \begin{align}
    \avr{r^2}_n^{\urm{Ca-44}}
    & =
      12.054 \pm 0.578
      \, \mathrm{fm}^2, \\
    \avr{r^2}_n^{\urm{Sn-110}}
    & =
      21.348 \pm 1.043
      \, \mathrm{fm}^2.
  \end{align}
\end{subequations}
Thus, the assumption for the neutron slope term is reasonable,
whereas the estimation of $ \avr{r^4}_p $ remains a problem.
\par
As an important remark, it should be noted that $ r_{\urm{E} n}^2 $ and $ r_{\urm{E} n}^4 $ include the information of the charge distribution of the neutron,
which has been determined with large uncertainty.
However, in the isotope-shift method proposed in this paper,
most of the contributions from these terms are canceled out in Eq.~\eqref{eq:master}.
In the last subsection of this section, discussion for uncertainty due to the nucleon form factors will be given.
\begin{table*}[tb]
  \centering
  \caption{Breakdown of uncertainties of isotope-shift method [Eq.~\eqref{eq:master}].
    Uncertainties are calculated by using standard deviations.
    For comparison, those of direct method is also shown.
    Uncertainties due to nucleon second and fourth moments
    $ r_{\urm{E} \tau}^2 $ and $ r_{\urm{E} \tau}^4 $
    (column with *) 
    are not considered in the total uncertainties.
    See the text for more details.}
  \label{tab:uncertainty}
  \begin{ruledtabular}
    \begin{tabular}{llddddddd}
      \multicolumn{1}{c}{\multirow{2}{*}{Nuclei}} & \multicolumn{1}{c}{\multirow{2}{*}{Method}}
      & \multicolumn{4}{c}{$ \sigma^2 $ ($ \mathrm{fm}^4 $)} & \multicolumn{1}{c}{\multirow{2}{*}{$ \sigma $ ($ \mathrm{fm}^2 $)}} & \multicolumn{2}{c}{$ \avr{r^2}_n $ ($ \mathrm{fm}^2 $)} \\
      \cline{3-6} \cline{8-9}
                                                  & & \multicolumn{1}{c}{$ r_{\urm{E} \tau}^n $ (*)} & \multicolumn{1}{c}{$ \avr{r^4}_p $} & \multicolumn{1}{c}{Neutron slope term} & \multicolumn{1}{c}{Total} & & \multicolumn{1}{c}{Calculation} & \multicolumn{1}{c}{Benchmark} \\
      \hline
      \multirow{2}{*}{$ \nuc{Ca}{44}{} $}
                                                  & Direct
      &  9.934 &    19.977 & \multicolumn{1}{c}{---}    &       19.977 &    4.470 & 12.842 & 12.234 \\
                                                  & Isotope-shift
      &  0.667 & 19566.142 & 0.334                      &    19566.476 &  139.880 & 14.639 & 12.234 \\
      \hline
      \multirow{2}{*}{$ \nuc{Sn}{110}{} $}
                                                  & Direct
      & 31.137 &     6961.160 & \multicolumn{1}{c}{---} &     6961.160 &   83.434 & 18.706 & 20.665 \\
                                                  & Isotope-shift
      &  4.091 & 15631629.395 & 1.087                   & 15631630.482 & 3953.686 & 28.774 & 20.665 \\ 
    \end{tabular}
  \end{ruledtabular}
\end{table*}
\subsection{Uncertainty due to nucleon form factors}
\label{sec:form_factor}
\par
Here, uncertainty due to nucleon form factors,
which is not considered in evaluations of $ \avr{r^2}_n $ in the previous subsections,
is discussed.
Note that, in this subsection, we do not consider the uncertainty discussed in the previous subsections.
\par
In general, $ \avr{r^2}_n $ can be regarded as a function of
$ r_{\urm{E} \tau}^2 $ and $ r_{\urm{E} \tau}^4 $.
Accordingly, the uncertainty due to the nucleon form factors can be calculated as
\begin{widetext}
  \begin{equation}
    \label{eq:uncertainty_formfactor}
    \sigma^2_{\urm{form}}
    \lesssim
    \left(
      \frac{\partial \avr{r^2}_n}{\partial r_{\urm{E} p}^2}
    \right)^2
    \sigma_{r_{\uurm{E} p}^2}^2
    +
    \left(
      \frac{\partial \avr{r^2}_n}{\partial r_{\urm{E} p}^4}
    \right)^2
    \sigma_{r_{\uurm{E} p}^4}^2
    +
    \left(
      \frac{\partial \avr{r^2}_n}{\partial r_{\urm{E} n}^2}
    \right)^2
    \sigma_{r_{\uurm{E} n}^2}^2
    +
    \left(
      \frac{\partial \avr{r^2}_n}{\partial r_{\urm{E} n}^4}
    \right)^2
    \sigma_{r_{\uurm{E} n}^4}^2,
  \end{equation}
  where contributions of the magnetic form factors are neglected since they are tiny.
  Here, contributions from the covariances are also neglected,
  and, because of this, the uncertainty is overestimated slightly.
  \par
  The uncertainty due to the nucleon form factors for the direct method [Eq.~\eqref{eq:fourth_rea}] can be estimated as
  \begin{equation}
    \label{eq:uncertainty_direct}
    \sigma_{\avr{r^2}_n}^2
    \lesssim 
    \left[
      \frac{Z}{N}
      \frac{1}{r_{\urm{E} n}^2}
      \left(
        \avr{r^2}_p
        -
        r_{\urm{E} p}^2
      \right)
    \right]^2
    \sigma_{r_{\uurm{E} p}^2}^2
    +
    \left(
      \frac{Z}{N}
      \frac{3}{10 r_{\urm{E} n}^2}
    \right)^2
    \sigma_{r_{\uurm{E} p}^4}^2
    +
    \left[
      \frac{1}{r_{\urm{E} n}^2}
      \left(
        \avr{r^2}_n
        -
        r_{\urm{E} p}^2
      \right)
    \right]^2
    \sigma_{r_{\uurm{E} n}^2}^2
    +
    \left(
      \frac{3}{10 r_{\urm{E} n}^2}
    \right)^2
    \sigma_{r_{\uurm{E} n}^4}^2 ,
  \end{equation}
  whereas the uncertainty due to the nucleon form factors for the isotope-shift method [Eq.~\eqref{eq:master}] can be estimated as
  \begin{align}
    \sigma_{\avr{r^2}_n}^2
    \lesssim & \,
               \left[
               1 -
               \frac{Z}{r_{\urm{E} n}^2}
               \frac{\avr{r^2}_{\urm{ch}}^{\left( N \right)} - \avr{r^2}_{\urm{ch}}^{\left( N - 2 \right)}}{2}
               -
               \frac{\kappa_n}{M_n^2}
               \frac{1}{r_{\urm{E} n}^2}
               \avr{\ve{l} \cdot \ve{\sigma}}_{n \urm{last}}
               \right]^2
               \sigma_{r_{\uurm{E} p}^2}^2
               \notag \\
             & \, 
               +
               \left[
               \frac{1}{r_{\urm{E} n}^2}
               \left(
               \avr{r^2}_n^{\left( N - 2 \right)}
               +
               N
               \frac{\avr{r^2}_n^{\left( N \right)} - \avr{r^2}_n^{\left( N - 2 \right)}}{2}
               -
               r_{\urm{E} p}^2
               \right)
               \right]^2
               \sigma_{r_{\uurm{E} n}^2}^2
               +
               \left(
               \frac{3}{10}
               \frac{1}{r_{\urm{E} n}^2}
               \right)^2
               \sigma_{r_{\uurm{E} n}^4}^2.
               \label{eq:uncertainty_isotope}
  \end{align}
\end{widetext}
For simplicity, here relative uncertainty $ \sigma_{r_{\uurm{E} \tau}^n} / r_{\urm{E} \tau}^n $, is assumed $ 5 \, \% $.
The uncertainties calculated by Eq.~\eqref{eq:uncertainty_direct}
for $ \nuc{Ca}{44}{} $ and $ \nuc{Sn}{110}{} $ 
are
$ \sigma^2_{\avr{r^2}_n} = 9.934 \, \mathrm{fm}^4 $
and $ 31.137 \, \mathrm{fm}^4 $, respectively.
If one uses the isotope-shift method,
the uncertainties are further suppressed as 
$ \sigma^2_{\avr{r^2}_n} = 0.667 \, \mathrm{fm}^4 $
and $ 4.091 \, \mathrm{fm}^4 $, respectively.
Thus, the isotope-shift method has another advantage to suppress the uncertainty due to the nucleon form factors.
These errors are anyway negligible as compared to the error introduced
by the correlation (that is, their covariance) between $ \avr{r^2}_p $ and $ \avr{r^4}_p $.

%% file: conclusion.tex
% -*- coding: utf-8 -*-
%%%%%%%%%%%%%%%%%%%%%%%%%%%%%%%%%%%%%%%%%%%%%%%%%% 
% 
% 4th moment paper manuscript
% Conclusion Part
% 
% Tomoya Naito, Gianluca Colo',
% Haozhao Liang, and Xavier Roca-Maza
% 
%%%%%%%%%%%%%%%%%%%%%%%%%%%%%%%%%%%%%%%%%%%%%%%%%%
%
\section{Conclusion}
\label{sec:conc}
\par
In this paper,
we have discussed
how to extract the neutron radius, that is, the second moment of the neutron distribution
by using the experimentally measured second and fourth moments of the charge distribution.
Our goal was to reduce model assumptions to a minimum.
To this aim, we have discussed in detail two contributions to the neutron moment:
the spin-orbit contribution and the contribution from the fourth moment of the proton distribution.
As for this latter, we have seen we can relate it to the second moment in a quite robust manner.
Therefore, we deem that we have been able to determine the mildest assumptions under which the neutron radius of a single isotope can be extracted.
\par
Our main result has been the introduction of a novel method to extract neutron radius from the charge density distribution
using the information of two neighboring even-even nuclei.
In this method, the uncertainties due to nucleon form factors and introduced by approximation for spin-orbit contribution are suppressed,
whereas the uncertainties introduced by the pairing are negligible.
We advocate that this method, namely the consideration of two neighboring isotopes, is more reliable.
\par
Despite our efforts, we conclude that the main obstacle to an accurate determination of the neutron radius is the contribution from the proton fourth moment.
Even if this is strongly correlated to the second moment, the resulting uncertainty cannot be neglected.
This uncertainty is strongly enhanced when propagated from the fourth moment to the neutron radius.
Eventually, 
extracting the neutron radius or the neutron-skin thickness
from the second and fourth moments of the charge density distribution
does not seem to be feasible based on the present discussion.
Despite these pessimistic conclusions,
the equations derived in this paper may be
useful for further understanding and investigation
and even more useful if, in the future, a clever way to better determine the proton fourth moment can be envisaged.

%% file: acknowledgement.tex
% -*- coding: utf-8 -*-
%%%%%%%%%%%%%%%%%%%%%%%%%%%%%%%%%%%%%%%%%%%%%%%%%% 
% 
% 4th moment paper manuscript
% Acknowledge Part
%
% Tomoya Naito, Gianluca Colo',
% Haozhao Liang, and Xavier Roca-Maza
% 
%%%%%%%%%%%%%%%%%%%%%%%%%%%%%%%%%%%%%%%%%%%%%%%%%%
%
\begin{acknowledgments}
  \par
  We would like to thank Y.~Guo, N.~Hinohara, and T.~Suda for fruitful discussions.
  T.N. and H.L. would like to thank the RIKEN iTHEMS program
  and the RIKEN Pioneering Project: Evolution of Matter in the Universe.
  T.N. acknowledges the JSPS Grant-in-Aid for JSPS Fellows under Grant No.~19J20543.
  H.L. acknowledges the JSPS Grant-in-Aid for Early-Career Scientists under Grant No.~18K13549
  and 
  the Grant-in-Aid for Scientific Research (S) under Grant No.~20H05648.
  G.C. and X.R.-M. acknowledge funding from the European Union's Horizon 2020 Research and Innovation Program under Grant No.~654002.
  The numerical calculations were performed on cluster computers at the RIKEN iTHEMS Program.
\end{acknowledgments}

%% file: appendix_detail.tex
% -*- coding: utf-8 -*-
%%%%%%%%%%%%%%%%%%%%%%%%%%%%%%%%%%%%%%%%%%%%%%%%%% 
% 
% 4th moment paper manuscript
% Appendix Part
% 
% Tomoya Naito, Gianluca Colo',
% Haozhao Liang, and Xavier Roca-Maza
% 
%%%%%%%%%%%%%%%%%%%%%%%%%%%%%%%%%%%%%%%%%%%%%%%%%% 
% 
\begin{widetext}
  \section{Derivation of Eq.~\eqref{eq:R2n_simple}}
  \label{sec:appen_detail}
  \par
  In this appendix, we recall the way to derive Eq.~\eqref{eq:R2n_simple},
  which was originally derived by Kurasawa and Suzuki~\cite{
    Kurasawa2019Prog.Theor.Exp.Phys.2019_113D01}. 
  Here, the unnormalized $ 2n $th moment of a density $ \rho $ can be calculated as 
  \begin{align}
    \int
    \rho \left( \ve{r} \right) \,
    r^{2n}
    \, d \ve{r}
    & =
      \int
      \left\{ 
      \frac{1}{\left( 2 \pi \right)^3}
      \int
      \left[
      \left( - \laplace_{\ve{q}} \right)^n
      \tilde{\rho} \left( \ve{q} \right)
      \right]
      e^{i \ve{q} \cdot \ve{r}}
      \, d \ve{q}
      \right\}
      d \ve{r}
      \notag \\
    & =
      \int
      \left[
      \left( - \laplace_{\ve{q}} \right)^n
      \tilde{\rho} \left( \ve{q} \right)
      \right]
      \left[
      \int
      \frac{1}{\left( 2 \pi \right)^3}
      e^{- i \left( - \ve{q} \right) \cdot \ve{r}}
      \, d \ve{r}
      \right]
      d \ve{q}
      \notag \\
    & =
      \int
      \left[
      \left( - \laplace_{\ve{q}} \right)^n
      \tilde{\rho} \left( \ve{q} \right)
      \right]
      \tilde{\delta} \left( - \ve{q} \right)
      \, d \ve{q}
      \notag \\
    & =
      \left[
      \left( - \laplace_{\ve{q}} \right)^n
      \tilde{\rho} \left( \ve{q} \right)
      \right]_{\ve{q} = \ve{0}},
      \label{eq:R2n_gen}
  \end{align}
  and hence
  \begin{align}
    \avr{r^{2n}}
    & =
      \frac{
      \left[
      \left( - \laplace_{\ve{q}} \right)^n
      \tilde{\rho} \left( \ve{q} \right)
      \right]_{\ve{q} = \ve{0}}}
      {\tilde{\rho} \left( \ve{0} \right)}
      \notag \\
    & =
      \frac{1}{N_{\tau}}
      \left[
      \left( - \laplace_{\ve{q}} \right)^n
      \tilde{\rho} \left( \ve{q} \right)
      \right]_{\ve{q} = \ve{0}},
      \label{eq:2nmoment}
  \end{align}
  holds~\cite{
    Kurasawa2019Prog.Theor.Exp.Phys.2019_113D01}.
  Since the Laplacian of a function $ f $ is written as
  \begin{align}
    \laplace f \left( r \right)
    & =
      \frac{d^2 f \left( r \right)}{dr^2}
      +
      \frac{2}{r} 
      \frac{d f \left( r \right)}{dr}
      \notag \\
    & =
      \frac{1}{r}
      \frac{d^2}{dr^2}
      \left[
      r f \left( r \right)
      \right], 
  \end{align}
  as long as $ f $ is spherically symmetric,
  Eq.~\eqref{eq:R2n_gen} can be simplified as
  \begin{align}
    \left[
    \left( - \laplace_{\ve{q}} \right)^n
    \tilde{\rho}_{\urm{ch}} \left( \ve{q} \right)
    \right]_{\ve{q} = \ve{0}}
    & =
      \left( -1 \right)^n
      \left\{
      \frac{1}{q}
      \frac{d^{2n}}{dq^{2n}}
      \left[
      q
      \sum_{\tau}
      \tilde{G}_{\urm{E} \tau} \left( q^2 \right)
      \tilde{\rho}_{\tau} \left( q \right)
      \right]
      \right\}_{q = 0}
      \notag \\
    & =
      \left( -1 \right)^n
      4 \pi 
      \left\{
      \frac{1}{q}
      \frac{d^{2n}}{dq^{2n}}
      \left[
      q
      \sum_{\tau}
      \tilde{G}_{\urm{E} \tau} \left( q^2 \right) \,
      \int_0^{\infty} 
      \rho_{\tau} \left( r \right) \,
      \frac{\sin \left( q r \right)}{qr}
      r^2 \, dr
      \right]
      \right\}_{q = 0}
      \notag \\
    & =
      \left( -1 \right)^n
      4 \pi 
      \sum_{\tau}
      \int_0^{\infty} 
      \left\{
      \frac{1}{q}
      \frac{d^{2n}}{dq^{2n}}
      \tilde{G}_{\urm{E} \tau} \left( q^2 \right) \,
      \sin \left( q r \right)
      \right\}_{q = 0}
      \rho_{\tau} \left( r \right) \,
      r \, dr.
      \label{eq:R2n_sph_2}
  \end{align}
  Thus, combining with Eq.~\eqref{eq:2nmoment} and assuming the spherical symmetry,
  we get Eq.~\eqref{eq:R2n_simple} for $ \rho_{\urm{ch}} $.
  \section{Charge form factor of the nucleus}
  \label{sec:appen_charge}
  \par
  In this appendix, 
  the detailed derivations of equations for the second and fourth moments discussed in Sec.~\ref{sec:general} are shown.
  The spin-orbit contributions to $ \avr{r^2}_{\urm{ch}} $ and $ \avr{r^4}_{\urm{ch}} $ are discussed as well.
  For the contribution to $ \avr{r^2}_{\urm{ch}} $, it has been already derived by
  Horowitz and Piekarewicz~\cite{
    Horowitz2012Phys.Rev.C86_045503}.
  Nevertheless, the contribution to $ \avr{r^4}_{\urm{ch}} $ can be derived in the parallel way as that to $ \avr{r^2}_{\urm{ch}} $,
  and, thus, the derivation of the former is also shown here for convenience.
  \par
  The Dirac and Pauli form factors of nucleons are denoted by
  $ \tilde{F}_{1 \tau} $ and $ \tilde{F}_{2 \tau} $,
  and the Sachs electric and magnetic form factors are defined by~\cite{
    Cloeet2009Few-BodySyst.46_1}
  \begin{subequations}
    \begin{align}
      \tilde{G}_{\urm{E} \tau} \left( Q^2 \right)
      & =
        \tilde{F}_{1 \tau} \left( Q^2 \right) - \frac{Q^2}{4 M_{\tau}^2} \tilde{F}_{2 \tau} \left( Q^2 \right), \\
      \tilde{G}_{\urm{M} \tau} \left( Q^2 \right)
      & =
        \tilde{F}_{1 \tau} \left( Q^2 \right) + \tilde{F}_{2 \tau} \left( Q^2 \right),
    \end{align}
  \end{subequations}
  respectively, in the relativistic scheme, where $ Q^2 = - q_{\mu} q^{\mu} > 0 $
  and $ M_{\tau} $ is the nucleon mass.
  The Dirac and Pauli form factors are normalized as
  \begin{subequations}
    \begin{align}
      \tilde{F}_{1 \tau} \left( 0 \right)
      & = e_{\tau}, \\
      \tilde{F}_{2 \tau} \left( 0 \right)
      & = \mu_{\tau},
    \end{align}
  \end{subequations}
  respectively,
  where $ e_{\tau} $ and $ \mu_{\tau} $ denote the charge and the magnetic dipole moment of nucleons $ \tau $.
  The anomalous magnetic moment $ \kappa_{\tau} $ reads
  $ \kappa_{\tau} = \mu_{\tau} - e_{\tau} $.
  Accordingly, the electric and magnetic form factors are normalized as
  \begin{subequations}
    \label{eq:form_factor_normalization}
    \begin{align}
      \tilde{G}_{\urm{E} p} \left( 0 \right)
      & =
        1, \\
      \tilde{G}_{\urm{E} n} \left( 0 \right)
      & =
        0, \\
      \tilde{G}_{\urm{M} \tau} \left( 0 \right)
      & =
        \mu_{\tau}.
    \end{align}
  \end{subequations}
  \par
  Throughout the derivation of the charge form factors, the relativistic scheme is used.
  The relativistic single-particle (Kohn-Sham) orbital under the spherically symmetric potential is written as
  \begin{equation}
    \phi_{a \tau} \left( \ve{r} \right)
    =
    \frac{1}{r}
    \begin{pmatrix}
      i g_{a \tau} \left( r \right) \\
      f_{a \tau} \left( r \right)
      \ve{\sigma} \cdot \hat{\ve{r}}
    \end{pmatrix}
    \ca{Y}_a \left( \theta, \phi \right)
    \chi_{\tau},
  \end{equation}
  where $ \hat{\ve{r}} = \ve{r} / r $,
  $ \ca{Y}_a $ is the spherical spinor,
  $ \chi_{\tau} $ is the isospin spinor,
  and the set of the quantum numbers $ a = \left( n, \kappa, m \right) $
  includes the principal quantum number $ n $,
  the angular quantum number,
  \begin{equation}
    \label{eq:kappa}
    \kappa
    =
    \begin{cases}
      + \left( j + 1/2 \right)
      & \text{(for $ j = l - 1/2 $)}, \\
      - \left( j + 1/2 \right)
      & \text{(for $ j = l + 1/2 $)},
    \end{cases}
  \end{equation}
  and its $ z $-projection $ m $ 
  \cite{
    Meng2006Prog.Part.Nucl.Phys.57_470,
    Niksic2011Prog.Part.Nucl.Phys.66_519,
    Liang2015Phys.Rep.570_1}.
  The normalization condition
  \begin{align}
    \int
    \phi_{a \tau}^{\dagger} \left( \ve{r} \right)
    \phi_{a \tau} \left( \ve{r} \right)
    \, d \ve{r}
    & =
      \int_0^{\infty}
      \left[
      \left\{
      g_{a \tau} \left( r \right)
      \right\}^2
      +
      \left\{
      f_{a \tau} \left( r \right)
      \right\}^2
      \right]
      \, dr
      \notag \\
    & = 1 
  \end{align}
  holds.
  \par
  The matrix element of the electromagnetic current operator for nucleon $ \tau $,
  $ \hat{J}^{\mu}_{\urm{EM} \tau} $,
  reads~\cite{
    ForestJr.1966Adv.Phys.15_1, 
    Donnelly1975Annu.Rev.Nucl.Sci.25_329,
    Friar1975AdvancesinNuclearPhysics8_219,
    Musolf1994Phys.Rep.239_1}
  \begin{equation}
    \hat{J}^{\mu}_{\urm{EM} \tau} \left( Q^2 \right) 
    =
    \tilde{G}_{\urm{E} \tau} \left( Q^2 \right) 
    \gamma^{\mu}
    +
    \frac{\tilde{G}_{\urm{M} \tau} \left( Q^2 \right) - \tilde{G}_{\urm{E} \tau} \left( Q^2 \right)}{1 + Q^2/4M_{\tau}^2}
    \left(
      \frac{Q^2}{4 M_{\tau}^2}
      \gamma^{\mu}
      +
      i \sigma^{\mu \nu}
      \frac{q_{\nu}}{2 M_{\tau}}
    \right)
  \end{equation}
  and especially
  its $ \mu = 0 $ component (density component) is 
  \begin{equation}
    \hat{J}^0_{\urm{EM} \tau} \left( Q^2 \right) 
    =
    \tilde{G}_{\urm{E} \tau} \left( Q^2 \right) 
    \gamma^0
    +
    \frac{\tilde{G}_{\urm{M} \tau} \left( Q^2 \right) - \tilde{G}_{\urm{E} \tau} \left( Q^2 \right)}{1 + Q^2/4 M_{\tau}^2}
    \left(
      \frac{Q^2}{4 M_{\tau}^2}
      \gamma^0
      +
      \gamma^0
      \frac{\ve{\gamma} \cdot \ve{q}}{2 M_{\tau}}
    \right).
  \end{equation}
  Since one-nucleon contribution of the form factors is defined by
  $ \brakket{N_{\tau} \left( p' \right)}
  {\hat{J}_{\urm{EM} \tau}^0 \left( Q^2 \right)}
  {N_{\tau} \left( p \right)} $ 
  with the on-shell nucleon state $ \ket{N_{\tau} \left( p \right)} $
  and four-momentum transfer $ Q = p' - p $,
  the charge form factor of the nucleus $ \tilde{\rho}_{\urm{ch}} $ is calculated
  by summing up the single-nucleon contributions as
  \begin{align}
    \tilde{\rho}_{\urm{ch}} \left( q \right)
    & \simeq
      \sum_{\text{occ}}
      \brakket{N_{\tau} \left( p' \right)}
      {\hat{J}_{\urm{EM} \tau}^0 \left( Q^2 \right)}
      {N_{\tau} \left( p \right)}
      \notag \\
    & =
      \sum_{\tau = p, \, n}
      \sum_a 
      \iint
      \phi_{a \tau}^{\dagger} \left( \ve{r}' \right)
      e^{i \ve{p}' \cdot \ve{r}'}
      \delta \left( \ve{r} - \ve{r}' \right) 
      {\hat{J}_{\urm{EM} \tau}^0 \left( Q^2 \right)}
      \phi_{a \tau} \left( \ve{r} \right)
      e^{- i \ve{p} \cdot \ve{r}}
      \, d \ve{r} \, d \ve{r}'
      \notag \\
    & \simeq
      \sum_{\tau = p, \, n}
      \sum_a 
      \iint
      \phi_{a \tau}^{\dagger} \left( \ve{r}' \right)
      e^{i \ve{p}' \cdot \ve{r}'}
      \delta \left( \ve{r} - \ve{r}' \right) 
      {\hat{J}_{\urm{EM} \tau}^0 \left( q^2 \right)}
      \phi_{a \tau} \left( \ve{r} \right)
      e^{- i \ve{p} \cdot \ve{r}}
      \, d \ve{r} \, d \ve{r}'
      \notag \\
    & =
      \sum_{\tau = p, \, n}
      \sum_a 
      \int
      \phi_{a \tau}^{\dagger} \left( \ve{r} \right)
      {\hat{J}_{\urm{EM} \tau}^0 \left( q^2 \right)}
      \phi_{a \tau} \left( \ve{r} \right)
      e^{- i \ve{q} \cdot \ve{r}}
      \, d \ve{r}
      \notag \\
    & =
      \sum_{\tau = p, \, n} 
      \left[
      \tilde{G}_{\urm{E} \tau} \left( q^2 \right)
      \tilde{F}_{\urm{V} \tau} \left( q \right)
      +
      \frac{\tilde{G}_{\urm{M}} \left( q^2 \right) - \tilde{G}_{\urm{E}} \left( q^2 \right)}{1 + q^2/4 M_{\tau}^2}
      \left(
      \frac{q^2}{4 M_{\tau}^2}
      \tilde{F}_{\urm{V} \tau} \left( q \right)
      +
      \frac{q}{2 M_{\tau}}
      \tilde{F}_{\urm{T} \tau} \left( q \right)
      \right)
      \right],
      \label{eq:appen_charge_1}
  \end{align}
  where $ \tilde{F}_{\urm{V} \tau} $ and $ \tilde{F}_{\urm{T} \tau} $ are the vector and tensor form factors
  defined by
  \begin{subequations}
    \begin{align}
      \tilde{F}_{\urm{V} \tau} \left( q \right) 
      & =
        \sum_a
        \int
        \phi_{a \tau}^{\dagger} \left( \ve{r} \right)
        \gamma^0
        \phi_{a \tau} \left( \ve{r} \right) 
        e^{i \ve{q} \cdot \ve{r}}
        \, d \ve{r},
        \label{eq:form_V1} \\
      \tilde{F}_{\urm{T} \tau} \left( q \right) 
      & =
        \sum_a 
        \int
        \phi_{a \tau}^{\dagger} \left( \ve{r} \right)
        \gamma^0
        \ve{\gamma} \cdot \hat{\ve{q}}
        \phi_{a \tau} \left( \ve{r} \right) 
        e^{i \ve{q} \cdot \ve{r}}
        \, d \ve{r},
        \label{eq:form_T1}
    \end{align}
  \end{subequations}
  respectively.
  The impulse approximation, which means that scattering occurs only once,
  and $ Q^2 = q^2 $ (elastic scattering, i.e., $ q^0 = 0 $) are introduced
  in $ {\simeq} $ appearing in the first and third lines of Eq.~\eqref{eq:appen_charge_1},
  respectively.
  Here, $ \tilde{\rho}_{\urm{ch}} $ is normalized to $ \tilde{\rho}_{\urm{ch}} \left( 0 \right) = Z $.
  Under the spherical symmetry, 
  Eqs.~\eqref{eq:form_V1} and \eqref{eq:form_T1} are calculated as 
  \begin{subequations}
    \begin{align}
      \tilde{F}_{\urm{V} \tau} \left( q \right) 
      = & \, 
          \sum_a
          \int
          \phi_{a \tau}^{\dagger} \left( r \right)
          \gamma^0 
          \phi_{a \tau} \left( r \right)
          j_0 \left(qr \right)
          \, d \ve{r}
          \notag \\
      = & \, 
          \sum_a
          \ca{N}_{a \tau}
          \int_0^{\infty}
          \left[
          \left\{
          g_{a \tau} \left( r \right)
          \right\}^2
          +
          \left\{
          f_{a \tau} \left( r \right)
          \right\}^2
          \right]
          j_0 \left(qr \right)
          \, dr,
          \label{eq:form_V2} \\
      \tilde{F}_{\urm{T} \tau} \left( q \right) 
      = & \,
          \sum_a
          \int
          \phi_{a \tau}^{\dagger} \left( r \right)
          \gamma^0
          \ve{\gamma} \cdot \hat{\ve{r}}
          \phi_{a \tau} \left( r \right)
          j_1 \left(qr \right)
          \, d \ve{r}
          \notag \\
      = & \,
          2
          \sum_a
          \ca{N}_{a \tau}
          \int_0^{\infty}
          g_{a \tau} \left( r \right)
          f_{a \tau} \left( r \right)
          j_1 \left(qr \right)
          \, dr,
          \label{eq:form_T2}
    \end{align}
  \end{subequations}
  where $ \ca{N}_{a \tau} $ is the occupation number of the orbital
  and 
  spherical Bessel functions,
  \begin{subequations}
    \begin{align}
      j_0 \left( r \right)
      & =
        \frac{\sin r}{r}, \\
      j_1 \left( r \right)
      & =
        \frac{\sin r}{r^2}
        - 
        \frac{\cos r}{r}
        =
        -
        \frac{d j_0 \left( r \right)}{dr}
    \end{align}
  \end{subequations}
  appear due to the spherical symmetry and the Fourier factor $ e^{i \ve{q} \cdot \ve{r}} $.
  \par
  The vector form factor $ \tilde{F}_{\urm{V} \tau} $ is the sum of the single-particle orbitals.
  Thus,
  $ \sum_a \ca{N}_{a \tau} \left[ \left\{ g_{a \tau} \left( r \right) \right\}^2 + \left\{ f_{a \tau} \left( r \right) \right\}^2 \right] $
  is the nucleon density distribution $ \rho_{\tau} $.
  Hence,
  \begin{equation}
    \tilde{F}_{\urm{V} \tau} \left( q \right)
    =
    \tilde{\rho}_{\tau} \left( q \right).
  \end{equation}
  \par
  In contrast, the tensor form factor $ \tilde{F}_{\urm{T} \tau} $ can be calculated as
  \begin{align}
    \frac{q}{2 M_{\tau}}
    \tilde{F}_{\urm{T} \tau} \left( q \right) 
    & = 
      \frac{2q}{2 M_{\tau}}
      \sum_a
      \ca{N}_{a \tau}
      \int_0^{\infty}
      g_{a \tau} \left( r \right)
      f_{a \tau} \left( r \right)
      j_1 \left(qr \right)
      \, dr
      \notag \\
    & \simeq
      \sum_a
      \ca{N}_{a \tau}
      \left[
      \frac{q^2}{3 M_{\tau}}
      \int_0^{\infty}
      g_{a \tau} \left( r \right)
      f_{a \tau} \left( r \right)
      r
      \, dr
      -
      \frac{q^4}{30 M_{\tau}}
      \int_0^{\infty}
      g_{a \tau} \left( r \right)
      f_{a \tau} \left( r \right)
      r^3
      \, dr
      \right]
  \end{align}
  because of
  \begin{equation}
    j_1 \left( q r \right)
    =
    \frac{rq}{3}
    - 
    \frac{r^3 q^3}{30}
    +
    O \left( r^5 q^5 \right).
  \end{equation}
  \par
  Using the Taylor expansion,
  \begin{equation}
    \label{eq:taylor_sinqr}
    \frac{\sin \left( q r \right)}{qr}
    =
    1
    -
    \frac{r^2 q^2}{6}
    +
    \frac{r^4 q^4}{120}
    +
    O \left( r^6 q^6 \right),
  \end{equation}
  and the normalization conditions shown in Eqs.~\eqref{eq:form_factor_normalization}
  \cite{
    Fuchs2004J.Phys.G30_1407},
  the contribution of nucleon $ \tau $ to the charge form factor $ \tilde{\rho}_{\urm{ch} \tau} $ is
  \begin{align}
    \tilde{\rho}_{\urm{ch} \tau} \left( q \right)
    \simeq & \, 
             \tilde{G}_{\urm{E} \tau} \left( q^2 \right)
             \tilde{F}_{\urm{V} \tau} \left( q \right)
             +
             \frac{\tilde{G}_{\urm{M} \tau} \left( q^2 \right) - \tilde{G}_{\urm{E} \tau} \left( q^2 \right)}{1 + q^2/4 M_{\tau}^2}
             \left(
             \frac{q^2}{4 M_{\tau}^2}
             \tilde{F}_{\urm{V} \tau} \left( q \right)
             +
             \frac{q}{2 M_{\tau}}
             \tilde{F}_{\urm{T} \tau} \left( q \right)
             \right)
             \notag \\
    \simeq & \,
             \left(
             e_{\tau} - \frac{q^2}{6} r_{\urm{E} \tau}^2 + \frac{q^4}{120} r_{\urm{E} \tau}^4
             \right)
             N_{\tau}
             \left(
             1 - \frac{q^2}{6} \avr{r^2}_{\tau} + \frac{q^4}{120} \avr{r^4}_{\tau}
             \right)
             \notag \\
           & \,
             +
             \left( 1 - \frac{q^2}{4 M_{\tau}^2} \right)
             \left[
             \left(
             \mu_{\tau} - \frac{q^2}{6} r_{\urm{M} \tau}^2
             \right)
             -
             \left(
             e_{\tau} - \frac{q^2}{6} r_{\urm{E} \tau}^2 
             \right)
             \right]
             \frac{q^2}{4 M_{\tau}^2}
             N_{\tau}
             \left(
             1 - \frac{q^2}{6} \avr{r^2}_{\tau} + \frac{q^4}{120} \avr{r^4}_{\tau}
             \right)
             \notag \\
           & \,
             +
             \left( 1 - \frac{q^2}{4 M_{\tau}^2} \right)
             \left[
             \left(
             \mu_{\tau} - \frac{q^2}{6} r_{\urm{M} \tau}^2
             \right)
             -
             \left(
             e_{\tau} - \frac{q^2}{6} r_{\urm{E} \tau}^2 
             \right)
             \right]
             \notag \\
           & \,
             \qquad
             \times
             \sum_a
             \ca{N}_{a \tau}
             \left[
             \frac{q^2}{3 M_{\tau}}
             \int_0^{\infty}
             r
             g_{a \tau} \left( r \right)
             f_{a \tau} \left( r \right)
             \, dr
             -
             \frac{q^4}{30 M_{\tau}}
             \int_0^{\infty}
             r^3
             g_{a \tau} \left( r \right)
             f_{a \tau} \left( r \right)
             \, dr
             \right]
             \notag \\
    = & \,
        \left(
        e_{\tau} - \frac{q^2}{6} r_{\urm{E} \tau}^2 + \frac{q^4}{120} r_{\urm{E} \tau}^4
        \right)
        N_{\tau}
        \left(
        1 - \frac{q^2}{6} \avr{r^2}_{\tau} + \frac{q^4}{120} \avr{r^4}_{\tau}
        \right)
        \notag \\
           & \,
             +
             \left( 1 - \frac{q^2}{4 M_{\tau}^2} \right)
             \left[
             \left(
             \mu_{\tau} - \frac{q^2}{6} r_{\urm{M} \tau}^2
             \right)
             -
             \left(
             e_{\tau} - \frac{q^2}{6} r_{\urm{E} \tau}^2 
             \right)
             \right]
             \frac{q^2}{4 M_{\tau}^2}
             N_{\tau}
             \left(
             1 - \frac{q^2}{6} \avr{r^2}_{\tau} + \frac{q^4}{120} \avr{r^4}_{\tau}
             +
             f_{\urm{T} 2 \tau}
             -
             q^2 
             f_{\urm{T} 4 \tau}
             \right)
             \notag \\
    \simeq & \,
             N_{\tau}
             e_{\tau}
             \notag \\
           & \, 
             -
             N_{\tau}
             q^2
             \left[
             \frac{1}{6}
             \left(
             e_{\tau} \avr{r^2}_{\tau}
             +
             r_{\urm{E} \tau}^2
             \right)
             -
             \frac{\kappa_{\tau}}{4 M_{\tau}^2}
             \left(
             1
             +
             f_{\urm{T} 2 \tau}
             \right)
             \right]
             \notag \\
           & \,
             +
             N_{\tau}
             q^4
             \left\{
             \frac{1}{360}
             \left(
             3 e_{\tau} \avr{r^4}_{\tau}
             +
             10 r_{\urm{E} \tau}^2 \avr{r^2}_{\tau}
             + 
             3 r_{\urm{E} \tau}^4
             \right)
             \right.
             \notag \\
           & \,
             \qquad
             \left.
             -
             \frac{1}{24 M_{\tau}^2}
             \left[
             \kappa_{\tau}
             \left(
             \avr{r^2}_{\tau}
             +
             6 f_{\urm{T} 4 \tau}
             \right) 
             +
             \left(
             r_{\urm{M} \tau}^2
             -
             r_{\urm{E} \tau}^2
             +
             \frac{3 \kappa_{\tau}}{2 M_{\tau}^2}
             \right)
             \left(
             1
             +
             f_{\urm{T} 2 \tau}
             \right)
             \right]
             \right\},
             \label{eq:appen_charge_expansion}
  \end{align}
  where
  $ e_p = 1 $ and $ e_n = 0 $ are the charge of protons and neutrons, respectively, 
  $ r_{\urm{E} \tau}^2 $ and $ r_{\urm{E} \tau}^4 $ is the second and fourth moments of nucleon $ \tau $, 
  $ r_{\urm{M} \tau}^2 $ are the second magnetic moment of nucleon $ \tau $,
  and
  \begin{subequations}
    \begin{align}
      f_{\urm{T} 2 \tau}
      & =
        \sum_a
        \ca{N}_{a \tau}
        \frac{4 M_{\tau}}{3 N_{\tau}}
        \int_0^{\infty}
        r
        g_{a \tau} \left( r \right)
        f_{a \tau} \left( r \right)
        \, dr, \\
      f_{\urm{T} 4 \tau}
      & =
        \sum_a
        \ca{N}_{a \tau}
        \frac{2 M_{\tau}}{15 N_{\tau}}
        \int_0^{\infty} 
        r^3
        g_{a \tau} \left( r \right)
        f_{a \tau} \left( r \right)
        \, dr.
    \end{align}
  \end{subequations}
  Therefore, comparing the same order of $ q^0 $, $ q^2 $, and $ q^4 $, we get
  \begin{subequations}
    \begin{align}
      Z
      = & \,
          Z,
          \label{eq:appen_zero_1} \\
      \avr{r^2}_{\urm{ch}}
      \simeq & \,
               \avr{r^2}_p
               +
               r_{\urm{E} p}^2
               +
               \frac{N}{Z}
               r_{\urm{E} n}^2
               -
               \frac{3}{2 M_{\tau}^2}
               \left[
               \kappa_p
               \left(
               1
               +
               f_{\urm{T} 2p}
               \right)
               +
               \frac{N}{Z}
               \kappa_n
               \left(
               1
               +
               f_{\urm{T} 2n}
               \right)
               \right], 
               \label{eq:appen_second_1} \\
      \avr{r^4}_{\urm{ch}}
      \simeq & \, 
               \left(
               \avr{r^4}_p
               +
               \frac{10}{3} r_{\urm{E} p}^2 \avr{r^2}_p
               + 
               r_{\urm{E} p}^4
               \right)
               +
               \frac{N}{Z}
               \left(
               \frac{10}{3} r_{\urm{E} n}^2 \avr{r^2}_n
               +
               r_{\urm{E} n}^4
               \right)
               \notag \\
        & \, 
          -
          \frac{5}{M_{\tau}^2}
          \left\{
          \kappa_p
          \left(
          \avr{r^2}_p
          +
          6 f_{\urm{T} 4p}
          \right) 
          +
          \left(
          r_{\urm{M} p}^2 - r_{\urm{E} p}^2
          +
          \frac{3 \kappa_p}{2 M_{\tau}^2}
          \right)
          \left(
          1
          +
          f_{\urm{T} 2p}
          \right)
          \right.
          \notag \\
        & \,
          \qquad
          \left.
          +
          \frac{N}{Z}
          \left[
          \kappa_n
          \left(
          \avr{r^2}_n
          +
          6 f_{\urm{T} 4n}
          \right) 
          +
          \left(
          r_{\urm{M} n}^2 - r_{\urm{E} n}^2
          +
          \frac{3 \kappa_n}{2 M_{\tau}^2}
          \right)
          \left(
          1
          +
          f_{\urm{T} 2n}
          \right)
          \right]
          \right\} ,
          \label{eq:appen_fourth_1}
    \end{align}
  \end{subequations}
  where the expansion of the charge form factor $ \tilde{\rho}_{\urm{ch}} $~\cite{
    Miller2019Phys.Rev.C99_035202},
  \begin{equation}
    \label{eq:taylor_charge}
    \tilde{\rho}_{\urm{ch}} \left( q \right)
    =
    Z
    \left[
      1
      -
      \frac{\avr{r^2}_{\urm{ch}}}{6}
      q^2
      + 
      \frac{\avr{r^4}_{\urm{ch}}}{120}
      q^4
      -
      \ldots
    \right]
  \end{equation}
  is used.
  \par
  In the free Dirac equation,
  \begin{equation}
    f_{a \tau} \left( r \right)
    =
    \frac{1}{2 M_{\tau}}
    \left(
      \frac{d}{dr}
      +
      \frac{\kappa}{r}
    \right)
    g_{a \tau} \left( r \right) 
  \end{equation}
  holds.
  Since the small component is small, this relationship is assumed to approximately hold even in the real nuclear systems.
  Using equations,
  \begin{subequations}
    \label{eq:appen_fTcalc_1}
    \begin{align}
      \int_0^{\infty}
      r
      g_{a \tau} \left( r \right)
      \frac{d}{dr}
      g_{a \tau} \left( r \right) 
      \, dr
      & = 
        \left[
        r
        g_{a \tau} \left( r \right)
        g_{a \tau} \left( r \right)
        \right]_0^{\infty}
        -
        \int_0^{\infty}
        \frac{d}{dr}
        \left[
        r
        g_{a \tau} \left( r \right)
        \right]
        g_{a \tau} \left( r \right) 
        \, dr
        \notag \\
      & = 
        -
        \int_0^{\infty}
        g_{a \tau} \left( r \right)
        g_{a \tau} \left( r \right) 
        \, dr 
        -
        \int_0^{\infty}
        r
        g_{a \tau} \left( r \right)
        \frac{d}{dr}
        g_{a \tau} \left( r \right) 
        \, dr
        \notag \\
      & = 
        -
        \frac{1}{2}
        \int_0^{\infty}
        g_{a \tau} \left( r \right)
        g_{a \tau} \left( r \right) 
        \, dr , \\
      \int_0^{\infty} 
      r^3
      g_{a \tau} \left( r \right)
      \frac{d}{dr}
      g_{a \tau} \left( r \right) 
      \, dr
      & = 
        \left[
        r^3
        g_{a \tau} \left( r \right)
        g_{a \tau} \left( r \right)
        \right]_0^{\infty}
        -
        \int_0^{\infty} 
        \frac{d}{dr}
        \left[
        r^3
        g_{a \tau} \left( r \right)
        \right]
        g_{a \tau} \left( r \right) 
        \, dr
        \notag \\
      & = 
        -
        \int_0^{\infty} 
        3 r^2
        g_{a \tau} \left( r \right)
        g_{a \tau} \left( r \right) 
        \, dr 
        -
        \int_0^{\infty} 
        r^3
        g_{a \tau} \left( r \right)
        \frac{d}{dr}
        g_{a \tau} \left( r \right) 
        \, dr
        \notag \\
      & = 
        -
        \frac{3}{2}
        \int_0^{\infty} 
        r^2
        g_{a \tau} \left( r \right)
        g_{a \tau} \left( r \right) 
        \, dr ,
    \end{align}
  \end{subequations}
  the integrals of $ f_{\urm{T} 2 \tau} $ and $ f_{\urm{T} 4 \tau} $ are
  \begin{subequations}
    \label{eq:appen_fTcalc_2}
    \begin{align}
      \int_0^{\infty}
      r
      g_{a \tau} \left( r \right)
      f_{a \tau} \left( r \right)
      \, dr
      & \simeq
        \frac{1}{2 M_{\tau}}
        \int_0^{\infty}
        r
        g_{a \tau} \left( r \right)
        \left(
        \frac{d}{dr}
        +
        \frac{\kappa}{r}
        \right)
        g_{a \tau} \left( r \right) 
        \, dr
        \notag \\
      & =
        -
        \frac{1}{2}
        \frac{1}{2 M_{\tau}}
        \int_0^{\infty}
        g_{a \tau} \left( r \right)
        g_{a \tau} \left( r \right) 
        \, dr 
        +
        \frac{\kappa}{2 M_{\tau}}
        \int_0^{\infty}
        g_{a \tau} \left( r \right)
        g_{a \tau} \left( r \right) 
        \, dr
        \notag \\
      & =
        \left( \kappa - \frac{1}{2} \right)
        \frac{1}{2 M_{\tau}} 
        \int_0^{\infty}
        g_{a \tau} \left( r \right)
        g_{a \tau} \left( r \right) 
        \, dr, \\
      \int_0^{\infty} 
      r^3
      g_{a \tau} \left( r \right)
      f_{a \tau} \left( r \right)
      \, dr
      & \simeq 
        \frac{1}{2 M_{\tau}}
        \int_0^{\infty} 
        r^3
        g_{a \tau} \left( r \right)
        \left(
        \frac{d}{dr}
        +
        \frac{\kappa}{r}
        \right)
        g_{a \tau} \left( r \right) 
        \, dr
        \notag \\
      & =
        -
        \frac{1}{2 M_{\tau}}
        \frac{3}{2}
        \int_0^{\infty} 
        r^2
        g_{a \tau} \left( r \right)
        g_{a \tau} \left( r \right) 
        \, dr 
        +
        \frac{\kappa}{2 M_{\tau}}
        \int_0^{\infty} 
        r^2
        g_{a \tau} \left( r \right)
        g_{a \tau} \left( r \right) 
        \, dr
        \notag \\
      & =
        \frac{1}{2 M_{\tau}}
        \left(
        \kappa
        -
        \frac{3}{2}
        \right)
        \int_0^{\infty} 
        r^2
        g_{a \tau} \left( r \right)
        g_{a \tau} \left( r \right) 
        \, dr.
    \end{align}
  \end{subequations}
  Therefore, parts of the spin-orbit contribution $ f_{\urm{T} 2 \tau} $ and $ f_{\urm{T} 4 \tau} $ are
  \begin{subequations}
    \label{eq:appen_fTcalc_3}
    \begin{align}
      f_{\urm{T} 2 \tau}
      & =
        \sum_a
        \ca{N}_{a \tau}
        \frac{4 M_{\tau}}{3 N_{\tau}}
        \int_0^{\infty}
        r
        g_{a \tau} \left( r \right)
        f_{a \tau} \left( r \right)
        \, dr
        \notag \\
      & = 
        \sum_a
        \ca{N}_{a \tau}
        \frac{4 M_{\tau}}{3 N_{\tau}}
        \left( \kappa - \frac{1}{2} \right)
        \frac{1}{2 M_{\tau}} 
        \int_0^{\infty}
        g_{a \tau} \left( r \right)
        g_{a \tau} \left( r \right) 
        \, dr
        \notag \\
      & \simeq
        \sum_a
        \ca{N}_{a \tau}
        \frac{2}{3 N_{\tau}}
        \left( \kappa - \frac{1}{2} \right), \\
      f_{\urm{T} 4 \tau}
      & =
        \sum_a
        \ca{N}_{a \tau}
        \frac{2 M_{\tau}}{15 N_{\tau}}
        \int_0^{\infty} 
        r^3
        g_{a \tau} \left( r \right)
        f_{a \tau} \left( r \right)
        \, dr
        \notag \\
      & =
        \sum_a
        \ca{N}_{a \tau}
        \frac{2 M_{\tau}}{15 N_{\tau}}
        \left(
        \kappa
        -
        \frac{3}{2}
        \right)
        \frac{1}{2 M_{\tau}}
        \int_0^{\infty} 
        r^2
        g_{a \tau} \left( r \right)
        g_{a \tau} \left( r \right) 
        \, dr
        \notag \\
      & \simeq
        \sum_a
        \ca{N}_{a \tau}
        \frac{1}{15 N_{\tau}}
        \left(
        \kappa
        -
        \frac{3}{2}
        \right)
        \avr{r^2}_{g_{a \tau}},
    \end{align}
  \end{subequations}
  where $ g_{a \tau} $ is assumed to be normalized and
  $ \avr{r^2}_{g_{a \tau}} $ is the second moment of $ g_{a \tau} $
  in the last lines of two equations above.
  Using
  \begin{align}
    2 \avr{\ve{l} \cdot \ve{s}}
    & =
      j \left( j + 1 \right)
      -
      l \left( l + 1 \right)
      -
      \frac{3}{4}
      \notag \\
    & =
      - \left( \kappa + 1 \right)
      \notag \\
    & =
      \begin{cases}
        - \left( l + 1 \right)
        & \text{($ \kappa > 0 $, i.e., $ j = l - 1/2 $)}, \\
        l
        & \text{($ \kappa < 0 $, i.e., $ j = l + 1/2 $)},
      \end{cases}
  \end{align}
  and
  assuming that each orbital is fully occupied or fully unoccupied,
  we get
  \begin{subequations}
    \label{eq:appen_fTcalc_4}
    \begin{align}
      1 + f_{\urm{T} 2 \tau}
      & \simeq
        \frac{2}{3 N_{\tau}}
        \left[
        \frac{3 N_{\tau}}{2}
        +
        \sum_a
        \left( 2j + 1 \right) 
        \left( \kappa - \frac{1}{2} \right)
        \right]
        \notag \\
      & =
        \frac{2}{3 N_{\tau}}
        \left[
        \frac{3 N_{\tau}}{2}
        -
        \sum_a
        \left( 2j + 1 \right)
        \left(
        2 \avr{\ve{l} \cdot \ve{s}}
        +
        \frac{3}{2}
        \right)
        \right]
        \notag \\
      & =
        -
        \frac{2}{3 N_{\tau}}
        \sum_a
        \left( 2j + 1 \right)
        2 \avr{\ve{l} \cdot \ve{s}}
        \notag \\
      & =
        \frac{4}{3 N_{\tau}}
        \sgn \left( \kappa \right)
        l \left( l + 1 \right) , \\
      \avr{r^2}_{\tau}^2 + 6 f_{\urm{T} 4 \tau}
      & \simeq
        \avr{r^2}_{\tau}^2
        +
        6
        \sum_a
        \left( 2j + 1 \right) 
        \frac{1}{15 N_{\tau}}
        \left(
        \kappa
        -
        \frac{3}{2}
        \right)
        \avr{r^2}_{g_{a \tau}}
        \notag \\
      & =
        \avr{r^2}_{\tau}^2
        -
        6
        \sum_a
        \left( 2j + 1 \right) 
        \frac{1}{15 N_{\tau}}
        \left(
        2 \avr{\ve{l} \cdot \ve{s}}
        +
        \frac{5}{2}
        \right)
        \avr{r^2}_{g_{a \tau}}
        \notag \\
      & =
        -
        \sum_a
        \left( 2j + 1 \right) 
        \frac{2}{5 N_{\tau}}
        2 \avr{\ve{l} \cdot \ve{s}}
        \avr{r^2}_{g_{a \tau}}.
    \end{align}
  \end{subequations}
  In total, the charge second and fourth moments are
  \begin{subequations}
    \begin{align}
      \avr{r^2}_{\urm{ch}}
      & \simeq 
        \avr{r^2}_p
        +
        r_{\urm{E} p}^2
        +
        \frac{N}{Z}
        r_{\urm{E} n}^2
        +
        \avr{r^2}_{\urm{SO} p}
        +
        \frac{N}{Z}
        \avr{r^2}_{\urm{SO} n},
        \label{eq:second_charge_all} \\
      \avr{r^4}_{\urm{ch}}
      & \simeq 
        \left(
        \avr{r^4}_p
        +
        \frac{10}{3} r_{\urm{E} p}^2 \avr{r^2}_p
        + 
        r_{\urm{E} p}^4
        \right)
        +
        \frac{N}{Z}
        \left(
        \frac{10}{3} r_{\urm{E} n}^2 \avr{r^2}_n
        +
        r_{\urm{E} n}^4
        \right)
        +
        \avr{r^4}_{\urm{SO} p}
        +
        \frac{N}{Z}
        \avr{r^4}_{\urm{SO} n},
        \label{eq:fourth_charge_all}
    \end{align}
  \end{subequations}
  where
  \begin{subequations}
    \begin{align}
      \avr{r^2}_{\urm{SO} \tau}
      & =
        -
        \frac{3 \kappa_{\tau}}{2 M_{\tau}^2}
        \left(
        1
        +
        f_{\urm{T} 2 \tau}
        \right)
        \notag \\
      & \simeq
        \frac{\kappa_{\tau}}{M_{\tau}^2 N_{\tau}}
        \sum_a
        \left( 2j + 1 \right)
        \avr{\ve{l} \cdot \ve{\sigma}},
        \label{eq:second_SO_appen} \\
      \avr{r^4}_{\urm{SO} \tau}
      & =
        - \frac{5}{M_{\tau}^2}
        \left[
        \kappa_{\tau}
        \left(
        \avr{r^2}_{\tau}^2
        +
        6 f_{\urm{T} 4 \tau}
        \right) 
        +
        \left(
        r_{\urm{M} \tau}^2
        -
        r_{\urm{E} \tau}^2
        +
        \frac{3 \kappa_{\tau}}{2 M_{\tau}^2}
        \right)
        \left(
        1
        +
        f_{\urm{T} 2 \tau}
        \right)
        \right]
        \notag \\
      & \simeq
        \frac{5}{M_{\tau}^2}
        \left[
        \kappa_{\tau}
        \sum_a
        \left( 2j + 1 \right) 
        \frac{2}{5 N_{\tau}}
        \avr{\ve{l} \cdot \ve{\sigma}}
        \avr{r^2}_{g_{a \tau}}
        +
        \left(
        r_{\urm{M} \tau}^2
        -
        r_{\urm{E} \tau}^2
        +
        \frac{3 \kappa_{\tau}}{2 M_{\tau}^2}
        \right)
        \frac{2}{3 N_{\tau}}
        \sum_a
        \left( 2j + 1 \right)
        \avr{\ve{l} \cdot \ve{\sigma}}
        \right]
        \notag \\
      & = 
        \frac{10}{M_{\tau}^2 N_{\tau}}
        \sum_a
        \left[
        \frac{\kappa_{\tau}}{5}
        \avr{r^2}_{g_{a \tau}}
        +
        \frac{r_{\urm{M} \tau}^2 - r_{\urm{E} \tau}^2}{3}
        +
        \frac{\kappa_{\tau}}{2 M_{\tau}^2}
        \right] 
        \left( 2j + 1 \right)
        \avr{\ve{l} \cdot \ve{\sigma}}.
        \label{eq:fourth_SO_appen}
    \end{align}
  \end{subequations}
  Note that Eq.~\eqref{eq:second_SO} was previously derived in Refs.~\cite{
    Chabanat1997Nucl.Phys.A627_710,
    Horowitz2012Phys.Rev.C86_045503},
  and if the orbital is not fully occupied, $ \left( 2j + 1 \right) $ is replaced by the occupation number $ \ca{N}_{a \tau} $.
  Also, contributions of the spin-orbit partners are canceled out if both orbitals are fully occupied.
  \par
  As long as only the electric form factors of nucleons are considered,
  the second and fourth moments are written as
  \begin{subequations}
    \begin{align}
      \avr{r^2}_{\urm{ch}}
      & \simeq
        \avr{r^2}_p
        +
        r_{\urm{E} p}^2 
        +
        \frac{N}{Z}
        r_{\urm{E} n}^2, \\
      \avr{r^4}_{\urm{ch}}
      & \simeq
        \avr{r^4}_p
        +
        \frac{10}{3}
        \left(
        r_{\urm{E} p}^2
        \avr{r^2}_p
        +
        \frac{N}{Z}
        r_{\urm{E} n}^2
        \avr{r^2}_n
        \right)
        +
        r_{\urm{E} p}^4
        +
        \frac{N}{Z}
        r_{\urm{E} n}^4.
    \end{align}
  \end{subequations}
  \section{Error estimation}
  \label{sec:appen_error}
  \par
  We note the error of $ f = X^{1/2} $ is estimated as $ \sigma_f^2 = \frac{1}{4X} \sigma_X^2 $,
  and thus
  $ \sigma_{\sqrt{\avr{r^2}}}^2 = \frac{1}{4 \avr{r^2}} \sigma_{\avr{r^2}}^2 $.
  The error of linear fitting $ \left( a \pm \Delta a \right) x + \left( b \pm \Delta b \right) $ is also estimated as
  $ \sigma^2 = \left( \Delta a \right)^2 x^2 + \left( \Delta b \right)^2 + 2 \rho_{ab} x \, \Delta a \, \Delta b $.
  Therefore, if there is perfect correlation ($ \rho_{ab} = \pm 1 $), 
  $ \sigma = \left| x \, \Delta a \pm \Delta b \right| $ holds.
\end{widetext}